\documentclass[aps,prd,showpacs]{revtex4}

\usepackage{amsmath}
\usepackage{amssymb}
\usepackage{bm}% bold math
\usepackage{graphicx}
\usepackage{multirow}
\usepackage{textcomp}

\allowdisplaybreaks[1]

\begin{document}

%\preprint{  }

\title{\boldmath Dispersive analysis of excited glueball states}
\author{Hsiang-nan Li}
\affiliation{Institute of Physics, Academia Sinica, Taipei 115}
%\affiliation{Institute of Physics, Academia Sinica,
%Taipei, Taiwan 115, Republic of China}

\date{\today}

\begin{abstract}

Motivated by the determination for the spin-parity quantum numbers of the $X(2370)$ meson 
at BESIII, we extend our dispersive analysis on hadronic ground states to excited states. 
The idea is to start with the dispersion relation which a correlation function obeys, 
and subtract the known ground-state contribution from the involved spectral density. 
Solving the resultant dispersion relation as an inverse problem with available
operator-product-expansion inputs, we extract excited-state masses from the 
subtracted spectral density. This formalism is verified by means of the application 
to the series of $\rho$ resonances, which establishes the $\rho(770)$, $\rho(1450)$ 
and $\rho(1700)$ mesons one by one under the sequential subtraction procedure. Our 
previous study has suggested the admixture of the $f_0(1370)$, $f_0(1500)$ and $f_0(1710)$ 
mesons (the $\eta(1760)$ meson) to be the lightest scalar (pseudoscalar) glueball. 
The present work predicts that the $f_0(2200)$ ($X(2370)$) meson is the first 
excited scalar (pseudoscalar) glueball.

\end{abstract}


\maketitle

\section{INTRODUCTION}

The quest for glueballs is an important mission in hadron physics, to which long-term 
efforts have been devoted. Though there is general consensus that the admixture of the 
$f_0(1370)$, $f_0(1500)$ and $f_0(1710)$ mesons is a convincing candidate for the 
lightest scalar glueball, the opinions on the candidate for the lightest pseudoscalar 
glueball remain diverse. Recently, the spin-parity quantum numbers and the mass of 
the $X(2370)$ meson were determined to be $0^{-+}$ and 
$2395\pm 11({\rm stat})^{+26}_{-94}({\rm syst})$ MeV, respectively, by the BESIII 
Collaboration \cite{BESIII:2023wfi} with the branching ratio 
${\rm Br}(J/\psi\to\gamma X(2370))\times {\rm Br}(X(2370)\to f_0(980)\eta')\times 
{\rm Br}(f_0(980)\to K_SK_S)=(1.31\pm 0.22({\rm stat})^{+2.85}_{-0.84}({\rm syst}))\times 10^{-5}$. 
That is, the production rate from the $J/\psi$ radiative decay is roughly 
${\rm Br}(J/\psi\to\gamma X(2370))\sim O(10^{-4})$, as 
${\rm Br}(X(2370)\to f_0(980)\eta')\times {\rm Br}(f_0(980)\to K_SK_S)\sim O(10^{-1})$ is 
assumed. It was then claimed that $X(2370)$ is the lightest pseudoscalar glueball in 
view of the similar mass range 2.3-2.6 GeV \cite{Bali,Morningstar:1999rf,Chen:2005mg,
Athenodorou:2020ani,Gui:2019dtm} and of the production rate $2.31(90)\times 10^{-4}$
\cite{Gui:2019dtm} predicted by quenched lattice QCD. However, it has been known that there 
are numerical and practical challenges in isolating glueballs from the relevant Euclidean 
correlators, when realistically light dynamical fermions are included 
\cite{Liu:2008ee,Cheng:2008ss}, particularly in the pseudoscalar-glueball case with the 
presence of the axial anomaly. According to \cite{Gui:2019dtm}, glueballs may have strong 
mixing with conventional mesons, such that quenched effects on glueball 
production rates cannot be reliably estimated either.

It is thus imperative to investigate the above subject in a different approach and find out 
whether alternative aspects can be attained. A widely adopted nonerturbative formalism is 
based on QCD sum rules \cite{SVZ}, which are constructed from a correlation function that 
defines a physical observable. Unfortunately, their application suffers significant 
theoretical uncertainties from the assumption of quark-hadron duality for parametrizing a 
spectral density and from the discretionary prescription for specifying a stability window in 
a Borel mass \cite{Coriano:1993yx,Coriano:1998ge,Leinweber:1995fn,Huang:1998wj,
Harnett:2000fy,Gubler:2010cf}. To improve this method, we proposed to handle QCD sum 
rules as an inverse problem \cite{Li:2020ejs}; the spectral density on the hadron side of a 
dispersion relation, including both resonance and continuum contributions, is regarded as an 
unknown in an integral equation, which is solved with standard operator-product-expansion (OPE) 
inputs on the quark side. The existence of a resonance is not presumed, the spectral density 
needs not be parametrized, a free continuum threshold is absent, the quark-hadron duality is 
not implemented, a Borel transformation is not required, and the discretionary prescription 
is not necessary, once a dispersion relation is solved directly. Furthermore, the precision 
of predictions can be enhanced systematically by adding higher-order and higher-power 
corrections to OPE inputs. The strength of the improved formalism compared to conventional 
sum rules has been elucidated in \cite{Li:2021gsx}.

Our approach, relying only on the analyticity of physical observables, has been applied to 
derivations of nonperturbative quantities associated with hadronic ground states, such as
masses \cite{Li:2020ejs,Li:2021gsx}, decay constants \cite{Li:2020ejs,Zhao:2024drr}
and the pion light-cone distribution amplitude \cite{Li:2022qul}. We demonstrated that a 
resonance peak shows up at the $\rho(770)$ meson mass in the spectral density naturally, as 
the dispersion relation for the correlator of vector quark currents is solved with the known
OPE inputs. The lightest scalar (pseudoscalar) glueball was predicted to be the admixture 
of the $f_0(1370)$, $f_0(1500)$ and $f_0(1710)$ mesons (the $\eta(1760)$ meson) in the same 
framework. Besides, the $f_0(500)$ meson (the admixture of the $\eta$ and $\eta'$ mesons), 
which is supposed to contain a small glue content, was uncovered together with the 
lightest scalar (pseudoscalar) glueball \cite{Li:2021gsx}. The new perspective of treating 
a dispersion relation as an inverse problem has been extended to the explanation of neutral 
meson mixing parameters \cite{Li:2020xrz,Xiong:2022uwj} and to the constraint on the hadronic 
vacuum-polarization contribution to the muon anomalous magnetic moment \cite{Li:2020fiz}. 
We will generalize the dispersive analysis on the ground states to excited states 
by subtracting the ground-state contributions from the spectral densities. The subtraction, 
attempting to suppress the interference between the excited and ground states, allows 
extraction of excited-state masses from the resultant dispersion relation.

The idea outlined above is first verified by establishing the series of $\rho$ resonances, 
i.e., the $\rho(770)$, $\rho(1450)$ and $\rho(1700)$ mesons one by one by under the 
repeated subtractions. The obtained masses 0.77, 1.47 and 1.65 GeV, respectively, 
are close to the observed values \cite{PDG}. More significant deviation from the data 
is noticed for higher states, which might be due to the propagation and accumulation
of uncertainties in the sequential subtraction procedure. Therefore, our exploration of 
excited $\rho$ mesons stops at $\rho(1700)$. After testing the method, we deduce 
straightforwardly the masses 2.17 (2.41) GeV for the first excited scalar (pseudoscalar) 
glueball. Confronting the results with \cite{PDG} and the BESIII measurement 
\cite{BESIII:2023wfi}, we tend to identify the acquired state as the $f_0(2200)$ ($X(2370)$) 
meson with the mass $2187\pm 14$ MeV ($2395\pm 11({\rm stat})^{+26}_{-94}({\rm syst})$ MeV). 
The model-dependent analysis in \cite{She:2024ewy} indicated that $X(2370)$ is a glueball-like 
particle. The production rate of $O(10^{-4})$ from the radiative decay 
$J/\psi\to\gamma f_0(2200)$ ($\gamma X(2370)$), lower than $O(10^{-3})$ associated with the 
ground state $f_0(1710)$ ($\eta(1760)$), seems reasonable. The search for $X(2370)$ at
the large hadron collider was assessed in \cite{Cao:2024mfn}. For recent studies on 
decays of the pseudoscalar glueball and its first excited state, refer to 
\cite{Eshraim:2019sgr,Eshraim:2020zrs}. 

%which assumed the masses 2.6 GeV and 
%3.7 GeV, following predictions from lattice QCD in the quenched approximation.
%The analysis was updated in \cite{Eshraim:2020zrs}, which lowed the ground-state 
%mass to that of $X(2370)$, motivated by the BESIII observation. 

The rest of the paper is organized as follows. In Sec.~II we illustrate how to solve for 
a spectral density from a dispersion relation with OPE inputs using the inverse matrix 
method \cite{Li:2021gsx}, and how to construct solutions for excited states via the 
subtraction procedure. It is certified that the solutions reveal the series of $\rho$ 
resonances successfully. In Sec.~III we probe the lightest and first excited 
glueballs in a similar way, and determine their masses. It is advocated that  
the $f_0(2200)$ and $X(2370)$ mesons are the potential candidates for the excited scalar 
and pseudoscalar glueballs, respectively. Section IV summarizes the outcomes in this work.

\section{SERIES OF $\rho$ RESONANCES}

\subsection{Dispersion Relation}

We briefly review the key ingredients of our theoretical setup. Consider the two-point correlator
\begin{eqnarray}
\Pi_{\mu\nu}(q^2)=i\int d^4xe^{iq\cdot x}
\langle 0|T[J_\mu(x)J_\nu(0)]|0\rangle=(q_\mu q_\nu-g_{\mu\nu}q^2)\Pi(q^2)\label{cur}
\end{eqnarray}
for the quark current $J_\mu=(\bar u\gamma_\mu u-\bar d\gamma_\mu d)/\sqrt{2}$.
The vacuum polarization function $\Pi(q^2)$ respects the identity 
\begin{eqnarray}
\frac{1}{2\pi i}\oint ds\frac{\Pi(s)}{s-q^2}=\Pi(q^2),\label{di1}
\end{eqnarray}
where the contour consists of two pieces of horizontal paths above and below the branch cut 
along the positive real axis on the complex $s$ plane, and a big circle $C$ of radius $R$ 
\cite{Li:2020ejs}. The OPE of the function $\Pi(q^2)$ in the deep Euclidean region of $q^2$ is 
reliable, so we have $\Pi^{\rm OPE}(q^2)$ \cite{SVZ} for the right-hand (quark) side of 
Eq.~(\ref{di1}),
\begin{eqnarray}
\Pi^{\rm OPE}(q^2)=\frac{1}{2\pi i}\oint ds\frac{\Pi^{\rm pert}(s)}{s-q^2}+
\frac{1}{12\pi}\frac{\langle\alpha_sG^2\rangle}{(q^2)^2}+
2\frac{\langle m_q \bar q q\rangle}{(q^2)^2} +\frac{224\pi}{81}
\frac{\kappa \alpha_s\langle \bar q q\rangle^2}{(q^2)^3},\label{ope}
\end{eqnarray}
up to the dimension-six condensate, ie., to the power correction of $1/(q^2)^3$. In 
Eq.~(\ref{ope}) the perturbative piece $\Pi^{\rm pert}(q^2)$ has been written as an integral 
over the same contour, 
$\langle\alpha_s G^2\rangle\equiv \langle \alpha_s G^a_{\mu\nu}G^{a\mu\nu}\rangle$ 
is the gluon condensate, $m_q$ is a light quark mass, and the parameter $\kappa=2$-4 
\cite{CDK,SN95,SN09} quantifies the violation in the factorization of the four-quark 
condensate $\langle (\bar q q)^2\rangle$ into the square of the quark condensate 
$\langle \bar q q\rangle$.

The contour integral on the left-hand (hadron) side of Eq.~(\ref{di1}) can be decomposed into
\begin{eqnarray}
\frac{1}{2\pi i}\oint ds\frac{\Pi(s)}{s-q^2}=
\frac{1}{\pi}\int_{0}^R ds\frac{{\rm Im}\Pi(s)}{s-q^2}
%+\frac{1}{\pi}\int_\Lambda^R ds\frac{{\rm Im}\Pi^{\rm pert}(s)}{s-q^2}
+\frac{1}{2\pi i}\int_C ds\frac{\Pi^{\rm pert}(s)}{s-q^2},\label{di2}
\end{eqnarray}
in which the imaginary part ${\rm Im}\Pi(s)$, involving nonperturbative dynamics from the small 
$s$ region of the branch cut, will be regarded as an unknown. The threshold, i.e., the lower 
bound of the dispersive integral, being of order of the pion mass squared, has been set to zero 
for simplicity. The integrand along the big circle $C$ has been replaced by 
$\Pi^{\rm pert}(s)$, because the perturbative calculation of $\Pi(s)$ is trustworthy for $s$ 
far away from physical regions, in accordance with the OPE in Eq.~(\ref{ope}). 
The equality of Eqs.~(\ref{ope}) and (\ref{di2}), following Eq.~(\ref{di1}), leads 
to the sum rule
\begin{eqnarray}
\frac{1}{\pi}\int_{0}^R ds\frac{{\rm Im}\Pi(s)}{s-q^2}=
\frac{1}{\pi}\int_{0}^R ds\frac{{\rm Im}\Pi^{\rm pert}(s)}{s-q^2}+
\frac{1}{12\pi}\frac{\langle\alpha_sG^2\rangle}{(q^2)^2}+
2\frac{\langle m_q \bar q q\rangle}{(q^2)^2} +\frac{224\pi}{81}
\frac{\kappa \alpha_s\langle \bar q q\rangle^2}{(q^2)^3},\label{di4}
\end{eqnarray}
where the contributions of $\Pi^{\rm pert}(s)$ from the big circle $C$ on the two sides
have canceled. The imaginary part of the perturbative piece is given by
\begin{eqnarray}
\frac{1}{\pi}{\rm Im}\Pi^{\rm pert}(s)=\frac{1}{4\pi^2}\left(1+\frac{\alpha_s}{\pi}\right)
\equiv c,\label{ope2}
%\Pi^{\rm pert}(q^2)&=&\frac{1}{4\pi^2}\left(1+\frac{\alpha_s}{\pi}\right)\ln\frac{\mu^2}{-q^2}
%\equiv c\ln\frac{\mu^2}{-q^2},
\end{eqnarray}
which defines the constant $c$ for later use.

We introduce a subtracted spectral density, related to the original one 
$\rho(s)\equiv {\rm Im}\Pi(s)/\pi$ via
\begin{eqnarray}
\Delta\rho(s,\Lambda)=\rho(s)-c[1-\exp(-s/\Lambda)].
\label{sub}
\end{eqnarray}
The arbitrary scale $\Lambda$ characterizes the transition of the unknown ${\rm Im}\Pi(s)$ 
to the perturbative expression ${\rm Im}\Pi^{\rm pert}(s)$. The smooth function 
$1-\exp(-s/\Lambda)$ decreases like $s$ at small $s$, and approaches the unity at large 
$s\gg\Lambda$, such that $\Delta\rho(s,\Lambda)$ retains the behavior of $\rho(s)\sim s$ near 
the threshold $s\to 0$ \cite{Kwon:2008vq}, and diminishes quickly as $s>\Lambda$. With the 
above features, the resonance structure of $\rho(s)$ in the small $s$ region is not altered 
by the subtraction term. We have confirmed that other smooth functions with the similar 
boundary and asymptotic behaviors yield basically identical solutions for $\rho(s)$. 
The radius $R$ in Eq.~(\ref{di4}) can be pushed toward the infinity, when the sum rule is 
formulated in terms of the subtracted spectral density,
\begin{eqnarray}
\int_{0}^\infty ds\frac{\Delta\rho(s,\Lambda)}{s-q^2}
&=&\int_{0}^\infty ds \frac{c e^{-s/\Lambda}}{s-q^2}
+\frac{1}{12\pi}\frac{\langle\alpha_sG^2\rangle}{(q^2)^2}+
2\frac{\langle m_q \bar q q\rangle}{(q^2)^2} +\frac{224\pi}{81}
\frac{\kappa \alpha_s\langle \bar q q\rangle^2}{(q^2)^3},\label{r20}
\end{eqnarray}
in which the dependence on the arbitrary radius $R$ has migrated to that on $\Lambda$.
We will solve for $\Delta\rho(s,\Lambda)$ under the two boundary conditions,
$\Delta\rho(s,\Lambda)\sim s$ at low $s$ and the fast diminishing of $\Delta\rho(s,\Lambda)$ 
at high $s$.

Since $\Delta\rho(s,\Lambda)$ is a dimensionless quantity, it can be cast into the form 
$\Delta\rho(s/\Lambda)$. Equation~(\ref{r20}) then becomes, under the variable changes 
$x= q^2/\Lambda$ and $y= s/\Lambda$,
\begin{eqnarray}
& &\int_{0}^\infty dy\frac{\Delta\rho(y)}{x-y}
=\int_{0}^\infty dy \frac{c e^{-y}}{x-y}
-\frac{1}{12\pi}\frac{\langle\alpha_sG^2\rangle}{x^2\Lambda^2}-
2\frac{\langle m_q \bar q q\rangle}{x^2\Lambda^2} -\frac{224\pi}{81}
\frac{\kappa \alpha_s\langle \bar q q\rangle^2}{x^3\Lambda^3},
\label{r21}
\end{eqnarray}
where $\Lambda$ in the subtracted spectral density has moved into the condensate terms to 
make them dimensionless. Note that the quark-hadron duality for the unknown spectral density 
is not assumed at any finite $y$ in the above derivation. The physical $\rho$ meson mass, if 
generated, corresponds to a peak location of $\Delta\rho(y)$. A physical solution should be
insensitive to the change of the arbitrary scale $\Lambda$, so a stability window in $\Lambda$ 
may show up, when $\Lambda$ increases from a low scale. As $\Lambda$ is further lifted, 
its effect disappears with the condensate, i.e., higher-power terms in Eq.~(\ref{r21}). Then 
a solution for $\Delta\rho(y)$ implies that a peak location of $\Delta\rho(s/\Lambda)$ in $s$ 
drifts with $\Lambda$. Once this scaling phenomenon occurs, none of the structure in 
$\Delta\rho(y)$ can be interpreted as a physical state. In the sense of searching for a 
stability window where a resonance mass stays constant, $\Lambda$ plays a role similar to a 
Borel mass in conventional sum rules.

\subsection{Inverse Matrix Method}

Equation~(\ref{r21}) is classified as the first kind of Fredholm integral equations, which 
has the typical form 
\begin{eqnarray}
\int_{0}^\infty dy\frac{\rho(y)}{x-y}= \omega(x),\label{su1}
\label{sum2}
\end{eqnarray}
with the unknown $\rho(y)$ and the input function $\omega(x)$. It will be solved in the 
inverse matrix method \cite{Li:2021gsx}, whose solid mathematical ground can be found in \cite{CWG}. 
Suppose that $\rho(y)$ decreases quickly enough with $y$, and the major contribution to the 
integral on the left-hand side comes from a finite range of $y$. It is then justified to 
expand the integral into a series in $1/x$ up to some power $N$ for a sufficiently large 
$|x|$ by inserting
\begin{eqnarray}
\frac{1}{x-y}=\sum_{m=1}^N \frac{y^{m-1}}{x^m}\label{ep1}
\end{eqnarray}
into Eq.~(\ref{su1}). Also suppose that $\omega(x)$ can be expanded into a power series in 
$1/x$ for a large $|x|$, 
\begin{eqnarray}
\omega(x)=\sum_{n=1}^N \frac{b_n}{x^n}.\label{bb1}
\end{eqnarray}

We expand the unknown into
\begin{eqnarray}
\rho(y)=\sum_{n=1}^N a_ny^\alpha e^{-y}L_{n-1}^{(\alpha)}(y)\label{r1}
\end{eqnarray}
using a set of generalized Laguerre functions $L_n^{(\alpha)}(y)$ up to degree $N-1$,
which satisfies the orthogonality
\begin{eqnarray}
\int_0^\infty y^\alpha e^{-y}L_m^{(\alpha)}(y)L_n^{(\alpha)}(y)dy
=\frac{\Gamma(n+\alpha+1)}{n!}\delta_{mn}.
\label{or1}
\end{eqnarray}  
The maximal integer $N$ will be fixed later, and the choice of the index $\alpha$ depends 
on the behavior of $\rho(y)$ around the boundary $y\to 0$. Substituting Eqs.~(\ref{ep1}), 
(\ref{bb1}) and (\ref{r1}) into Eq.~(\ref{su1}), and equating the coefficients of $1/x^n$
(recall that $|x|$ is still arbitrary despite of being large), we arrive at the matrix 
equation $Ma=b$ with the matrix elements
\begin{eqnarray}
M_{mn}=\int_{0}^\infty dy y^{m-1+\alpha}e^{-y}L_{n-1}^{(\alpha)}(y),\label{m2}
\end{eqnarray}
the unknown vector $a=(a_1, a_2,\cdots,a_N)$ and the input vector $b=(b_1,b_2,\cdots,b_N)$, 
where $m$ and $n$ run from 1 to $N$. 

If the inverse matrix $M^{-1}$ exists, one can get a solution for $a$ from the known $b$ through 
$a=M^{-1}b$ trivially. The existence of $M^{-1}$ thus infers the uniqueness of the solution. 
In principle, the true solution is approached by increasing the number 
of polynomials $N$ in Eq.~(\ref{r1}). The distinction between the true solution and the 
approximate one causes a small power correction $1/x^{N+1}$ to the left-hand side of 
Eq.~(\ref{su1}) owing to the orthogonality in Eq.~(\ref{or1}). The orthogonality also 
demands $M_{mn}=0$ for $m<n$. Namely, $M$ is a triangular matrix, so the coefficients 
$a_n$ built up previously are not amended, when an additional higher-degree polynomial is 
included into the expansion in Eq.~(\ref{r1}). Nevertheless, both $m$ and $n$ have to stop 
at a finite $N$ in a practical application, for the determinant of $M$ diminishes with 
its dimension. An approximate solution for $a$ would then deviate from the true 
solution violently, when a tiny fluctuation of the input $b$ is amplified by the 
huge elements of $M^{-1}$. A consequence of the violent deviation may be reflected by, for 
instance, the lose of the positivity of a spectral density. This is a generic feature of an 
ill-posed inverse problem. Hence, the optimal choice of $N$ is set to either the integer
corresponding to the minimal $a_N$, for which a solution is stable against the variation of 
$N$, or to the maximal integer, above which the positivity of a spectral density is violated.

\subsection{Ground State}

We begin with the extraction of the ground-state $\rho(770)$ meson mass $m_\rho$ from
Eq.~(\ref{r21}). The boundary condition of $\Delta\rho(y)\sim y$ at $y\to 0$ stated before
designates the index $\alpha=1$ in Eq.~(\ref{r1}). The matrix elements $M_{mn}$ are 
prepared according to Eq.~(\ref{m2}). The input coefficients $b_n$ are computed following 
the right-hand side of Eq.~(\ref{r21}), among which $b_2$ and $b_3$ of the $1/x^2$ and $1/x^3$ 
terms, respectively, receive additional contributions from the condensates. The OPE parameters 
and the strong coupling $\alpha_s$, evaluated at the scale of 1 GeV, are chosen as
\cite{CDK,SN95,SN09,Wang:2016sdt,Narison:2014wqa,SVZ,SN98,Forkel:2003mk},
\begin{eqnarray}
& &\langle m_q\bar q q\rangle = 0.007\times(-0.246)^3\;{\rm GeV}^4,\;\;
\langle\alpha_sG^2\rangle=0.08\; {\rm GeV}^4,\nonumber\\
& &\alpha_s\langle \bar q q\rangle^2 = 1.49\times 10^{-4}\;{\rm GeV}^6,\;\;
\alpha_s=0.5,\;\; \kappa=2.5.\label{put}
\end{eqnarray}
The above set of parameters are basically the same as in our previous analysis 
\cite{Li:2021gsx}, except the minor difference of the factorization violation parameter $\kappa$.
We do not aim at a precise accommodation of the measured value $m_\rho=775.26\pm 0.23$ MeV 
\cite{PDG}, but at yielding $m_\rho\approx 0.77$ GeV. It has been examined that the 
renormalization-group evolutions of $\alpha_s$ and the condensates around the scales 1-2 GeV 
have a negligible impact on results of $m_\rho$ \cite{Li:2021gsx}. We do not consider 
higher-order corrections to the condensate pieces \cite{Wang:2016sdt,ST90}, but vary the
gluon condensate $\langle\alpha_sG^2\rangle$, i.e, the $1/x^2$ term, by 20\% \cite{Li:2021gsx}
to mimic the uncertainty from the OPE inputs.

%for their effects can be mimicked by tuning $\kappa$. 

\begin{figure}
\begin{center}
\includegraphics[scale=0.28]{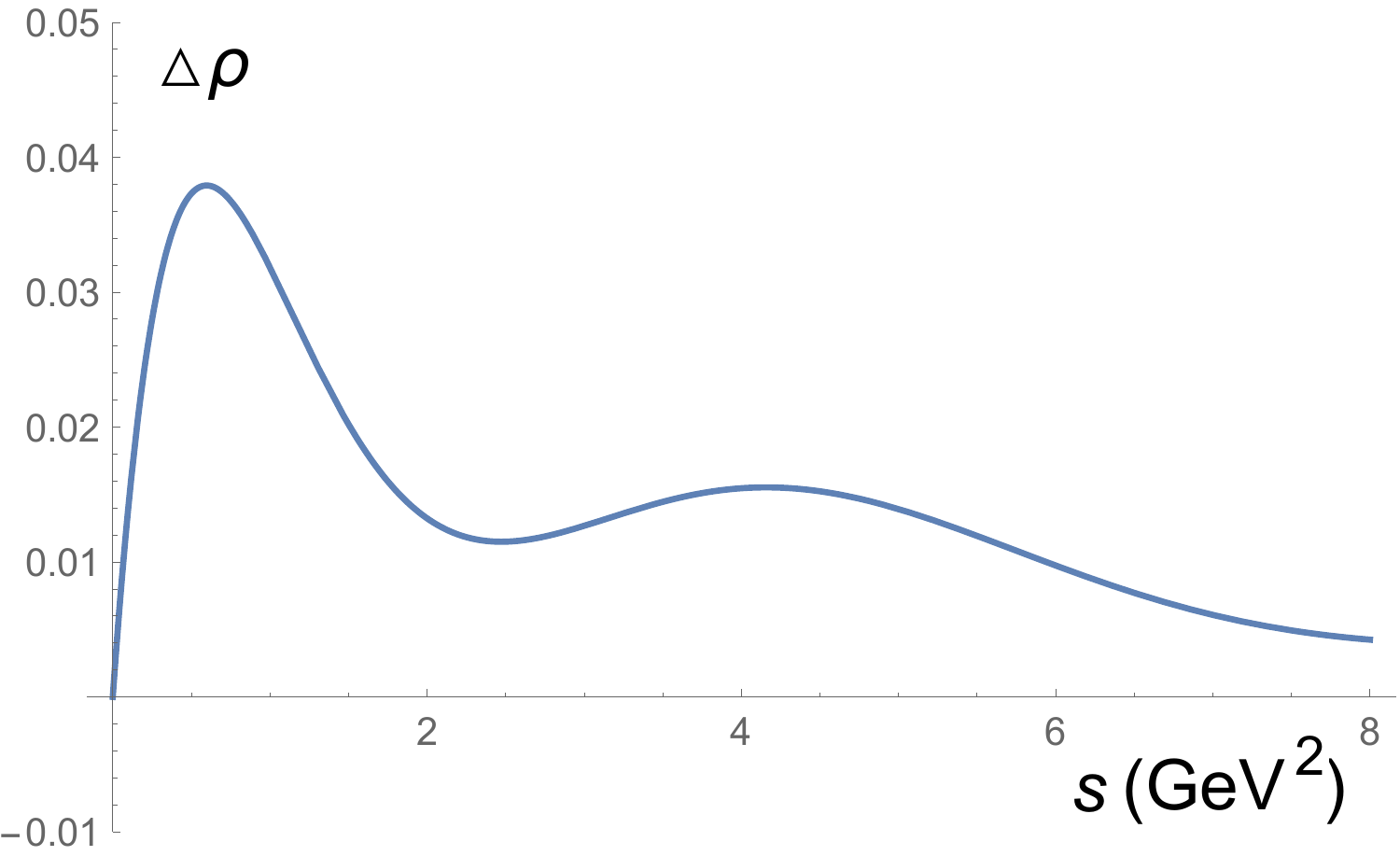}\hspace{1.0cm}
\includegraphics[scale=0.28]{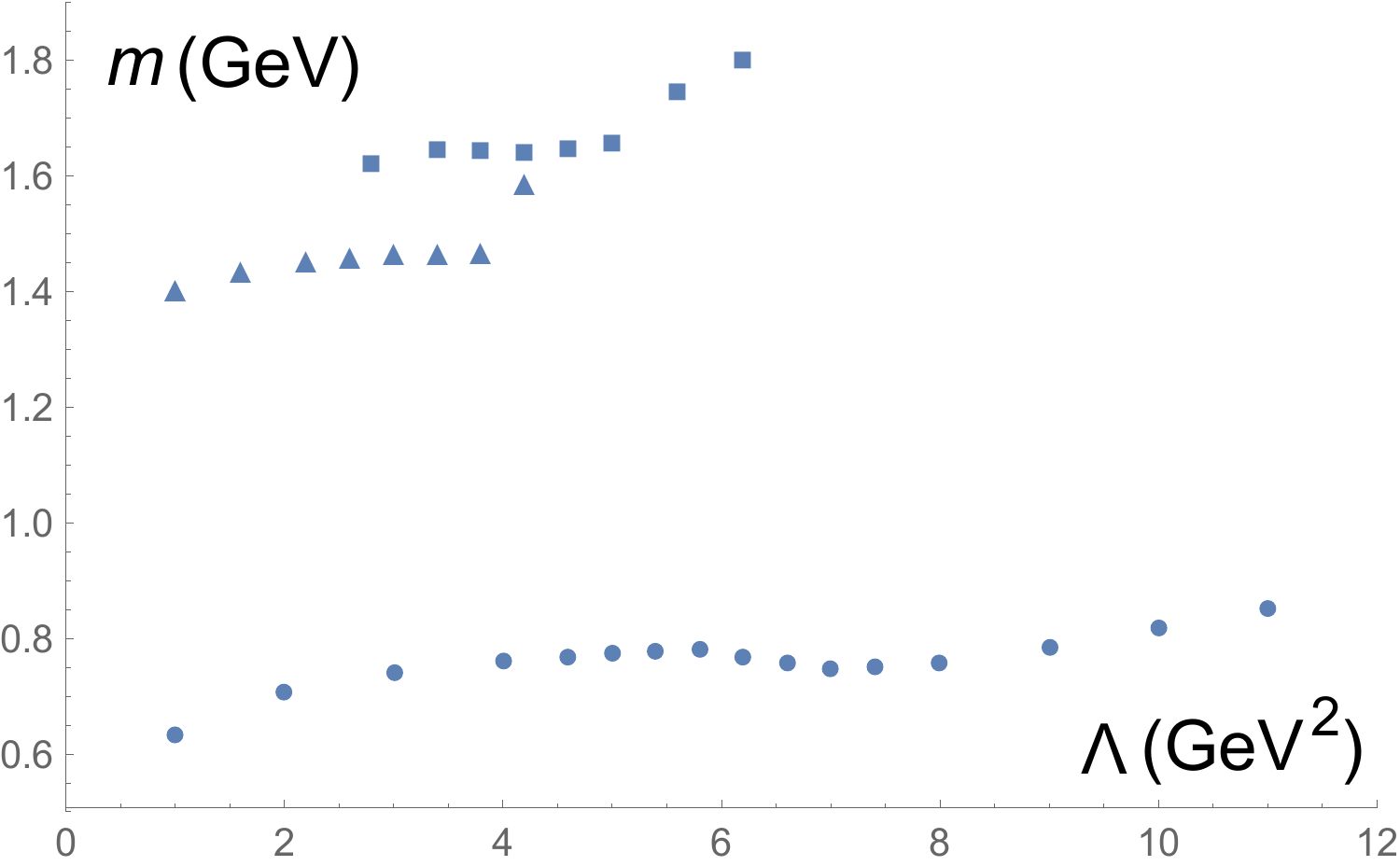}

(a) \hspace{7.0 cm} (b)
\caption{\label{fig1}
(a) $s$ dependence of the ground-state solution $\Delta \rho_0(s,\Lambda)$ for 
$\Lambda=5.0$ GeV$^2$. (b) $\Lambda$ dependencies of $m_\rho$ (circles), $m_{\rho'}$ (triangles)
and $m_{\rho''}$ (squares), where the stability windows are spotted by denser markers.}
\end{center}
\end{figure}

We derive the inverse matrix $M^{-1}$, the unknowns $a_n$ from the given $b_n$, and the 
ground-state solution 
$\Delta \rho_0(s,\Lambda)=(s/\Lambda)\exp(-s/\Lambda)\sum_{n=1}^N a_n L_{n-1}^{(1)}(s/\Lambda)$ 
as an expansion in the generalized Laguerre polynomials. The outcomes of 
$\Delta \rho_0(s,\Lambda)$ for the scale $\Lambda=5.0$ GeV$^2$ with $N=32$ and 33 are 
indistinguishable, assuring the stability of the solutions against the variation of $N$; the 
small ratios of the coefficients, $|a_{33}/a_{1}|\approx 10^{-3}$ and 
$|a_{33}/a_{32}|\approx 0.2$, guarantee the excellent convergence of the polynomial expansion.
The matrix elements of $M^{-1}$ increase rapidly as $N>33$, manifested by the ratio 
$a_{34}/a_{33}\approx 6$. The positivity of the solved spectral density 
$\rho_0(s)=\Delta \rho_0(s,\Lambda)+c[1-\exp(-s/\Lambda)]$ is substantiated in this case. We 
read off the $\rho(770)$ meson mass $m_\rho=0.77$ GeV from the location $s=0.60$ GeV$^2$ of 
the major peak of $\Delta \rho_0(s,\Lambda)$ for $\Lambda=5.0$ GeV$^2$ and $N=33$ displayed in 
Fig.~\ref{fig1}(a). The oscillatory tail at $s>2$ GeV$^2$ indicates that the continuum 
contribution to the spectral density $\rho_0(s)$ must differ from the perturbative one 
$c= 0.029$. In other words, the local quark-hadron duality does not hold actually. It should 
be pointed out that $m_\rho$ extracted above is smaller than from the peak location of the 
spectral density $\rho_0(s)$ \cite{Li:2021gsx} by 4\%, which is not crucial. We switch to the
subtracted spectral density for determining resonance masses, because this option is more 
practical for glueball analyses as demonstrated in the next section.

%and consistent with those obtained in the the maximum entropy method \cite{Ohtani:2012ps}

We then solve for the subtracted spectral densities from Eq.~(\ref{r21}) with various 
$\Lambda$, and obtain the $\rho(700)$ meson masses $m_\rho$ from their peak locations. The 
degree $N$, at which $a_N$ reaches a minimum, increases with $\Lambda$, taking the integers 
8, 14, 20, 26, 33, 40 and 48 for $\Lambda=1.0$, 2.0, 3.0, 4.0. 5.0, 6.0 and 7.0 GeV$^2$ and 
beyond, respectively. The $\Lambda$ dependence of $m_\rho$ within 
$1.0\le\Lambda \le 11.0$ GeV$^2$ is depicted in Fig.~\ref{fig1}(b), where the expected features 
are salient: the curve climbs from $\Lambda=1.0$ GeV$^2$, goes up and down mildly between 
$m_\rho=(0.77\pm 0.02)$ GeV in the interval $4.0\le\Lambda \le 8.0$ GeV$^2$, and then ascends 
monotonically with $\Lambda$ for $\Lambda> 8.0$ GeV$^2$. As postulated before, a physical 
resonance should be insensitive to the arbitrary scale $\Lambda$, so the stability 
window $4.0\le\Lambda \le 8.0$ GeV$^2$ puts forth the result $m_\rho\approx 0.77$ GeV. The 
growth of $m_\rho$ at $\Lambda>8.0$ GeV$^2$ is attributed to the disappearance of the 
condensate effects. The location of the broad bump in Fig.~\ref{fig1}(a), shifting with
$\Lambda$, does not meet the stability criterion, such that it cannot be interpreted as a 
physical state. The ground-state mass depends weakly on the OPE parameters 
\cite{Li:2021gsx}; the representative variation of the gluon 
condensate $\langle\alpha_sG^2\rangle$ by $\pm 20\%$ changes $m_\rho$ by $\mp 5\%$. Combining 
the theoretical uncertainties from the fluctuation in the stability window and from the 
OPE inputs, we conclude our prediction for the $\rho(770)$
meson mass $m_\rho=(0.77\pm 0.04)$ GeV in consistency with the data \cite{PDG}.

\subsection{Excited States}

To access an excited state, the contribution of a ground state needs to be deducted from the 
correlator, i.e., from the spectral density in order to suppress the interference between 
them. For this purpose, we employ the strategy in \cite{Li:2020ejs}, 
parametrizing the $\rho(770)$ contribution as a $\delta$-function $F_0\delta(s-m_\rho^2)$, 
and subtracting it from the two sides of the dispersion relation in Eq.~(\ref{r20}).  
The profile of the spectral density for a resonance is described by a Breit-Wigner formula, 
so the $\delta$-function parametrization can be viewed as a narrow-width approximation. 
Besides, it has been noticed that the maximal degree $N$ in the polynomial expansion 
increases with the scale $\Lambda$. To reduce the $1/(q^2)^{N+1}$ power correction 
resulting from the truncation in the polynomial expansion, we pick up the ground-state solution 
$\Delta\rho_0(s,\Lambda)$ corresponding to the high end of the stability window, i.e., 
$\Lambda=8.0$ GeV$^2$ for evaluating the strength of the $\delta$-function,
\begin{eqnarray}
F_0=\int_0^\infty ds\Delta\rho_0(s,\Lambda)=0.22\;{\rm GeV}^2.\label{sf0}
\end{eqnarray}
We have affirmed that another choice of $\Delta\rho_0(s,\Lambda)$ corresponding to a smaller 
$\Lambda=7.4$ GeV$^2$ does not affect the outcome for the first excited state. The unknown on 
the left-hand side of Eq.~(\ref{r20}) after the ground-state subtraction can be renamed as 
$\Delta\rho(s)$. The associated dispersion relation is then expressed, under the variable changes 
$x=q^2/\Lambda$ and $y=s/\Lambda$, as
\begin{eqnarray}
\int_{0}^\infty dy\frac{\Delta\rho(y)}{x-y}
=\int_{0}^\infty dy \frac{c e^{-y}-f_0\delta(y-r_0)}{x-y}
-\frac{1}{12\pi}\frac{\langle\alpha_sG^2\rangle}{x^2\Lambda^2}-
2\frac{\langle m_q \bar q q\rangle}{x^2\Lambda^2} -\frac{224\pi}{81}
\frac{\kappa \alpha_s\langle \bar q q\rangle^2}{x^3\Lambda^3},
\label{r22}
\end{eqnarray}
with the ratios $f_0=F_0/\Lambda$ and $r_0=m_\rho^2/\Lambda$. The perturbative piece in the 
OPE has been modified by the subtraction explicitly. 

%According to the orthogonality condition in Eq.~(\ref{or1}), $F_0=a_1\Lambda$ is 
%proportional to the first coefficient $a_1$ in the polynomial expansion in Eq.~(\ref{r1}), 
%which is not altered by the variation of $N$.

\begin{figure}
\begin{center}
\includegraphics[scale=0.28]{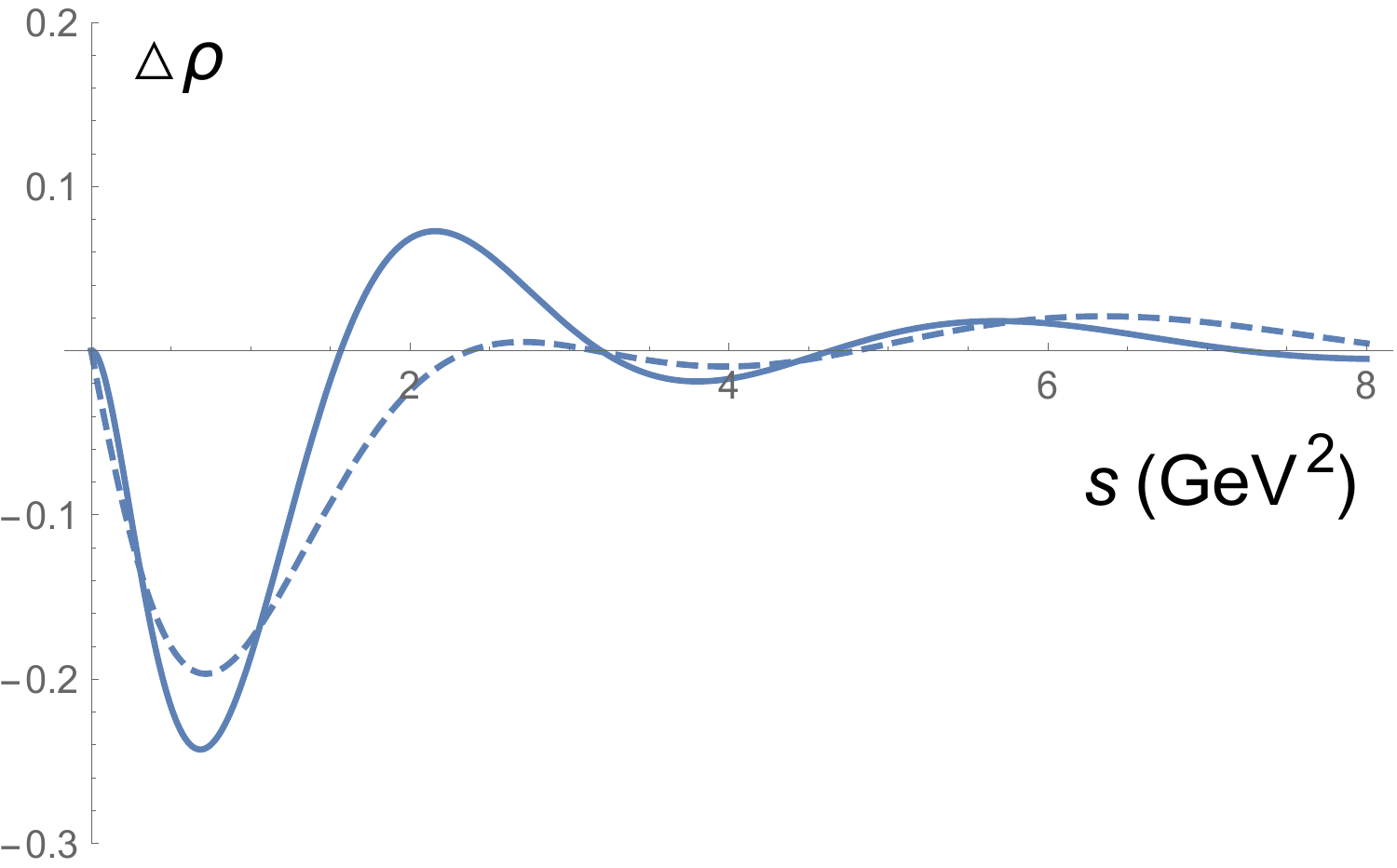}\hspace{1.0cm}
\includegraphics[scale=0.28]{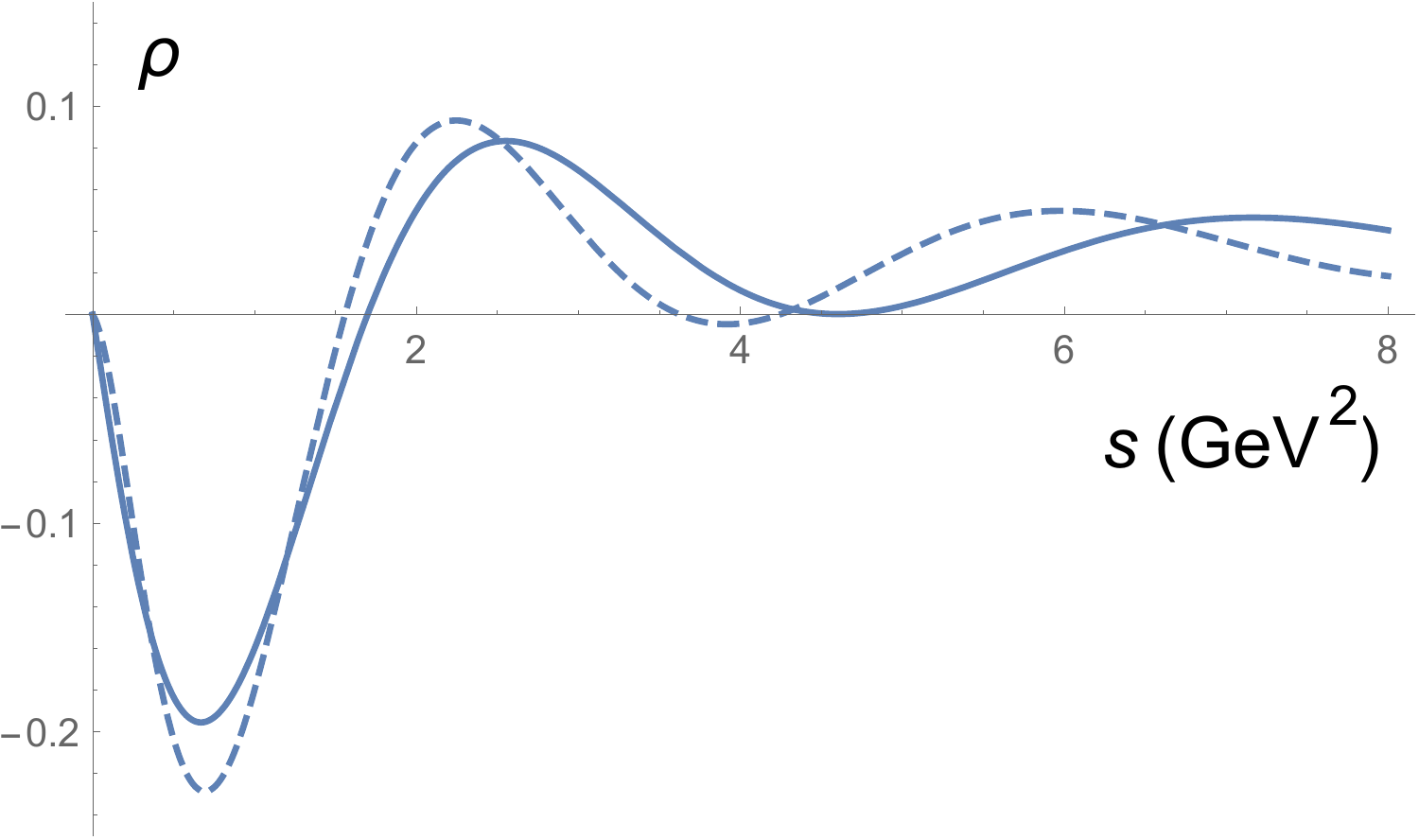}

(a) \hspace{7.0 cm} (b)
\caption{\label{fig2}
(a) $s$ dependencies of the first excited-state solution $\Delta\rho_1(s,\Lambda)$ for  
$\Lambda=3.0$ GeV$^2$ (solid line) and of the second excited-state solution 
$\Delta\rho_2(s,\Lambda)$ for $\Lambda=4.6$ GeV$^2$ (dashed line). (b) $s$ dependencies 
of the solutions $\rho_1(s)=\Delta \rho_1(s,\Lambda)+c[1-\exp(-s/\Lambda)]$ for $\Lambda=4.2$ 
GeV$^2$ with $N=38$ (solide line) and $N=49$ (dashed line).}
\end{center}
\end{figure}

The input coefficients $b_n$ arise from the right-hand side of Eq.~(\ref{r22}) with the 
identical parameters in Eq.~(\ref{put}). Strictly speaking, the strong coupling constant 
$\alpha_s$ should decrease a bit, since a higher energy region will be probed. Nevertheless, 
the running effect has been found to be negligible \cite{Li:2021gsx}. The same matrix 
elements $M_{mn}$ in Eq.~(\ref{m2}) lead to the unknowns $a_n$ and the solution  
$\Delta\rho_1(s,\Lambda)$ for the subtracted spectral density. The behavior of 
$\Delta\rho_1(s,\Lambda)$ in $s$ for $\Lambda=3.0$ GeV$^2$ with the maximal degree $N=40$ is 
presented in Fig.~\ref{fig2}(a). The ratios of the coefficients, $|a_{40}/a_1|\approx 10^{-2}$ 
and $|a_{40}/a_{39}|\approx 0.3$, imply the satisfactory convergence of $a_n$ and the 
stability of the solution against the variation of $N$. The deep valley around $s= 0.6$ 
GeV$^2$ originates from the subtraction of the $\rho(770)$ contribution. The major peak, 
located at $s=2.16$ GeV$^2$, i.e., at the first excited-state mass $m_{\rho'}=1.47$ GeV,  
matches well the observed $\rho(1450)$ meson mass $1465\pm 25$ MeV \cite{PDG}. The oscillation 
above $s=2$ GeV$^2$, like that in Fig.~\ref{fig1}(a), is tied to higher continuum 
contributions. It has been remarked \cite{PDG} that there may exist a nearby state $\rho(1570)$
of mass $1570\pm 36 \pm 62$ MeV, which is speculated to be due to an 
Okubo-Zweig-Iizuka-suppressed decay mode of $\rho(1700)$. The $\rho(1450)$ and $\rho(1570)$ 
states cannot be resolved in our setup; more precise inputs are needed for resolving finer 
structures in a spectrum as having been elaborated in \cite{Li:2021gsx}.

We explore the dependence of the mass $m_{\rho'}$ on the scale $\Lambda$  
by scanning the interval $1.0\le\Lambda\le 4.2$ GeV$^2$. The maximal degree $N$, at which
the coefficient $a_N$ reaches a minimum, takes the integers 14, 22, 30, 35, 40, 45 and 
48 for $\Lambda=1.0$, 1.6, 2.2, 2.6, 3.0, 3.4 and 3.8 GeV$^2$, respectively. As $\Lambda$
is greater than 4.0 GeV$^2$, the positivity requirement for a spectral density is activated, 
and we have to terminate the polynomial expansion before $a_N$ reaches a minimum. For instance, 
the highest degree must be set to $N=38$ for $\Lambda=4.2$ GeV$^2$, though $a_{49}$ is the 
minimal coefficient. Figure~\ref{fig2}(b) compares the two solutions 
$\rho_1(s)=\Delta \rho_1(s,\Lambda)+c[1-\exp(-s/\Lambda)]$ for the spectral density
at $\Lambda=4.2$ GeV$^2$ with $N=38$ and $N=49$. Ignoring the valleys from the ground-state 
subtraction around $s=0.6$ GeV$^2$, we see that the $N=38$ curve remains positive for 
$s\gtrsim 2$ GeV$^2$, while the $N=49$ curve turns negative at $s\approx 4.0$ GeV$^2$. The 
$\Lambda$ dependence of $m_{\rho'}$ in Fig.~\ref{fig1}(b) ascends from $\Lambda=1.0$ GeV$^2$, 
and exhibits a plateau of the height $m_{\rho'}\approx 1.47$ GeV in the stability window 
$2.2\le \Lambda\le 3.8$ GeV$^2$, which is narrower than in the $\rho(770)$ case. The abrupt 
rise of the curve for $\Lambda > 3.8$ GeV$^2$ is induced by the aforementioned positivity 
constraint, and signals the entering into the scaling regime. The physical solutions for the 
subtracted spectral density within the stability window give the $\rho(1450)$ meson mass 
$m_{\rho'}=1.47\pm 0.01$ GeV. The $\pm 20\%$ variation of the gluon condensate generates only 
about $\mp 1\%$ change on the results of $m_{\rho'}$. Combining the two sources of 
uncertainties, we provide our prediction $m_{\rho'}=1.47\pm 0.02$ GeV in good agreement with 
the data \cite{PDG}. 

We should bear in mind that there are other potential sources of theoretical uncertainties; 
if the ground-state contribution is parametrized by a Breit-Wigner type function with the 
width $\Gamma_\rho\approx 0.15$ GeV \cite{PDG} and the normalization being fixed by 
Eq.~(\ref{sf0}), $m_{\rho'}$ will increase by 5\%, an amount similar to that from the 
variation of the gluon condensate. However, once a finite-width parametrization is adopted, 
the interference between the ground and excited states, such as the phase difference between 
them, becomes more nontrivial. The subtraction procedure then needs to be managed by 
taking into account the interference effects in a delicate way, such that the 5\% enhancement 
may be modified. Obviously, the formulation of these new ingredients goes beyond the scope of 
the present work, so we will stick to the simple $\delta$-function ansatz.

For the extension to the second excited state $\rho(1700)$, we implement the
narrow-width approximation again, parametrizing the profile of the $\rho(1450)$
contribution to the spectral density as a $\delta$-function $F_1\delta(s-m_{\rho'}^2)$. 
The solution $\Delta\rho_1(s,\Lambda)$ corresponding to the high end of the stability window, 
i.e., $\Lambda=3.8$ GeV$^2$ with $N=48$, is selected for the evaluation of the strength
\begin{eqnarray}
F_1=\int_{t_1}^\infty ds\Delta\rho_1(s)=0.11\;{\rm GeV}^2.\label{t1}
\end{eqnarray}
The threshold $t_1=1.56$ GeV$^2$ is set to the first root in $s>0$ of the equation
$\Delta\rho_1(s,\Lambda)=0$. Namely, the deep valley attributed to the ground-state subtraction
is discarded from the above integral. We admit that there exists ambiguity in choosing the
threshold for the $\rho(1450)$ profile because of the narrow-width approximation already
made for the $\rho(770)$ profile. As shown shortly, this ambiguity does render the outcome 
for the $\rho(1700)$ meson mass more distinct from the measured value.

The dispersion relation for the second excited state is written as
\begin{eqnarray}
\int_{0}^\infty dy\frac{\Delta\rho(y)}{x-y}
=\int_{0}^\infty dy \frac{c e^{-y}-f_0\delta(y-r_0)-f_1\delta(y-r_1)}{x-y}
-\frac{1}{12\pi}\frac{\langle\alpha_sG^2\rangle}{x^2\Lambda^2}-
2\frac{\langle m_q \bar q q\rangle}{x^2\Lambda^2} -\frac{224\pi}{81}
\frac{\kappa \alpha_s\langle \bar q q\rangle^2}{x^3\Lambda^3},
\label{r23}
\end{eqnarray}
with the ratios $f_1=F_1/\Lambda$ and $r_1=m_{\rho'}^2/\Lambda$, where the contributions of 
the lowest two resonances have been removed from the perutrbative piece. 
We deduce the solution $\Delta\rho_2(s,\Lambda)$ from Eq.~(\ref{r23}) in a similar manner, 
whose $s$ dependence for $\Lambda=4.6$ GeV$^2$ with the degree $N=45$ is plotted in 
Fig.~\ref{fig2}(a). Here $N=45$ is demanded by the positivity constraint on the 
spectral density, instead of by the minimal coefficient $a_N$. The ratios of the coefficients, 
$|a_{45}/a_1|\approx 0.07$ and $|a_{45}/a_{44}|\approx 1$, indicate that the polynomial
expansion is still under control. The deep valley, resulting from the subtraction of the 
$\rho(770)$ and $\rho(1450)$ contributions, is a bit broader than in the previous case with 
only the ground-state subtraction. Note that the first peak next to the valley, located at 
$s=2.71$ GeV$^2$, i.e., at the second excited-state mass $m_{\rho''}= 1.65$ GeV, is shorter 
than the bump at larger $s\approx 6.0$ GeV$^2$, in contrast to the shape of the $\rho(1450)$ 
curve. It warns of the reliability of this solution, a concern which has been raised before.

To investigate the dependence of the mass $m_{\rho''}$ on the 
scale $\Lambda$, we gather the solutions $\Delta\rho_2(s,\Lambda)$ in the interval 
$2.8\le\Lambda\le 6.2$ GeV$^2$. The positivity constraint is effective at lower $\Lambda$ till
$\Lambda=5.0$ GeV$^2$. The highest degree $N$ takes the integers 29, 34, 38, 42 and 45 for 
$\Lambda=2.8$, 3.4, 3.8, 4.2 and 4.6 GeV$^2$, respectively. Then $N$ assumes 49, 49 and
48 for $\Lambda=5.0$, 5.6 and 6.2 GeV$^2$, respectively, at which the coefficient $a_N$ is
minimal. The quality for the convergence of the polynomial expansion is acceptable for all 
the above solutions. The $\Lambda$ dependence of $m_{\rho''}$ in Fig.~\ref{fig1}(b) exhibits 
a plateau of the height $m_{\rho''}\approx 1.65$ GeV in the stability window 
$3.4\le \Lambda\le 5.0$ GeV$^2$, which is also narrower than in the $\rho(770)$ case. We 
read off $m_{\rho''}=1.65\pm 0.01$ GeV from Fig.~\ref{fig1}(b), that falls below the observed 
$\rho(1700)$ meson mass $1720\pm 20$ MeV \cite{PDG}. The $20\%$ variation of the gluon
condensate has little impact on this prediction. The deviation is 
understandable, since the uncertainties involved in the determinations for lower states 
will be transported into those for higher states and enlarged through the sequential 
subtractions. The consistency with the data can be improved by decreasing the 
threshold $t_1$ in Eq.~(\ref{t1}) slightly. However, we do not attempt such tuning, and
stop pursuing higher states.

\section{EXCITED STATES OF GLUEBALLS}

We are ready to dive into the more ambiguous topic on the excited states of scalar and 
pseudoscalar glueballs. The advantage of our method for extracting
glueball properties has been summarized in \cite{Li:2021gsx}. Other theoretical approaches to 
glueball physics were introduced in the review \cite{Mathieu:2008me}; appropriate moments of 
a Borel transformed correlator have to be chosen to form ratios for the extraction of a 
glueball mass in sum rules, because the stability window in a Borel mass may not exist for 
ratios of other moments \cite{SN98,Huang:1998wj}. On the other hand, the lower (higher) moments 
are more sensitive to light (heavy) resonances \cite{SN98}. The above subtlety is not an
issue to our formalism, in which full information of the OPE inputs is utilized to solve a 
dispersion relation. We will demonstrate that definite predictions for the masses of the
excited scalar and pseudoscalar glueballs can be made.

\subsection{Dispersion Relation}

Define the correlation function for the glueball channel
\begin{eqnarray}
\Pi_G(q^2)=i\int d^4xe^{iq\cdot x}\langle 0|TO_G(x)O_G(0)|0\rangle,\label{c1}
\end{eqnarray}
$G=S$ or $P$, where the local composite operators $O_S(x)=\alpha_s G^a_{\mu\nu}(x)G^{a\mu\nu}(x)$ 
and $O_P(x)=\alpha_s G^a_{\mu\nu}(x){\tilde G}^{a\mu\nu}(x)$ denote the gluonic interpolating 
fields for the scalar $(0^{++})$ and pseudoscalar $(0^{-+})$ glueballs, respectively, 
$\tilde G_{\mu\nu}\equiv i\epsilon_{\mu\nu\rho\sigma}G^{\rho\sigma}/2$ 
being the dual of the gluon field strength. We have the OPE of the correlation function 
$\Pi_G(q^2)$ in the deep Euclidean region of $q^2$ \cite{SVZ},
\begin{eqnarray}
\Pi_G^{\rm OPE}(q^2)&=&q^4 \ln\frac{-q^2}{\mu^2}\left[A_0^{(G)}+A_1^{(G)}\ln\frac{-q^2}{\mu^2}
+A_2^{(G)}\ln^2\frac{-q^2}{\mu^2}\right]\nonumber\\
& &+\left[B_0^{(G)}+B_1^{(G)} \ln\frac{-q^2}{\mu^2}\right]\langle \alpha_s G^2\rangle
-\left[C_0^{(G)}+C_1^{(G)} \ln\frac{-q^2}{\mu^2}\right]\frac{\langle g G^3\rangle}{q^2}+
D_0^{(G)}\frac{\langle \alpha_s^2 G^4\rangle_G}{(q^2)^2},\label{dig3}
\end{eqnarray}
up to the dimension-eight condensate, ie., up to the power correction of $1/(q^2)^2$,
$\mu$ being a renormalization scale. Although the instanton background may contribute to 
the OPE \cite{NSVZ,Forkel:2003mk,Forkel:2000fd,Harnett:2000fy,Wen:2010as,Wen:2010qoe,Wang:2015mla}, 
we do not include its effect as in \cite{SN98} to avoid the model dependence from, e.g., 
parametrizations for instanton size distributions. The minor quark-loop and quark-condensate 
corrections \cite{Yuan:2009vs} are ignored too.

The coefficients in Eq.~(\ref{dig3}) for the scalar glueball 
are \cite{NSVZ79,Kataev:1981gr,BS90,CKS97,HS01}
\begin{eqnarray}
A_0^{(S)}& =&-2\left(\frac{\alpha_s}{\pi}\right)^2\left[1 +
\frac{659}{36}\frac{\alpha_s}{\pi}+ 247.48\left(\frac{\alpha_s}{\pi}\right)^2\right],\nonumber\\
A_1^{(S)}& =& 2\left(\frac{\alpha_s}{\pi}\right)^3\left(\frac{\beta_0}{4} +
65.781\frac{\alpha_s}{\pi}\right),\;\;\;\;
A_2^{(S)}=-10.125\left(\frac{\alpha_s}{\pi}\right)^4,\nonumber\\
B_0^{(S)}&=& 4\alpha_s\left(1 +\frac{175}{36}\frac{\alpha_s}{\pi}\right),\;\;\;\;
B_1^{(S)}= -\frac{\alpha_s^2}{\pi}\beta_0,\nonumber\\
C_0^{(S)}&=& 8\alpha_s^2,\;\;\;\;
C_1^{(S)}= 0,\;\;\;\;
D_0^{(S)}= 8\pi\alpha_s,
\end{eqnarray}
with the factor $\beta_0 = 11 - 2n_f/3=9$, $n_f$ being the number of active quarks.
Those for the pseudoscalar glueball are \cite{NSVZ79,ZS03,AMM92}
\begin{eqnarray}
A_0^{(P)}& =& -2\left(\frac{\alpha_s}{\pi}\right)^2\left[1 +
20.750\frac{\alpha_s}{\pi}+ 305.95\left(\frac{\alpha_s}{\pi}\right)^2\right],\nonumber\\
A_1^{(P)}& =& 2\left(\frac{\alpha_s}{\pi}\right)^3\left(\frac{\beta_0}{4} +
72.531\frac{\alpha_s}{\pi}\right),\;\;\;\;
A_2^{(P)}=-10.125\left(\frac{\alpha_s}{\pi}\right)^4,\nonumber\\
B_0^{(P)}&=& 4\alpha_s,\;\;\;\;
B_1^{(P)}= \frac{\alpha_s^2}{\pi}\beta_0\nonumber\\
C_0^{(P)}&=& -8\alpha_s^2,\;\;\;\;
C_1^{(P)}= 0,\;\;\;\;
D_0^{(P)}= 4\pi\alpha_s.\label{ap}
\end{eqnarray}

Following Eqs.~(\ref{di1})-(\ref{di4}), we establish the dispersion relation
for the correlator in Eq.~(\ref{c1}), 
\begin{eqnarray}
\frac{1}{\pi}\int_{0}^R ds\frac{{\rm Im}\Pi_G(s)}{s-q^2}
&=& \frac{1}{\pi}\int_{0}^R ds\frac{{\rm Im}\Pi_G^{\rm pert}(s)}{s-q^2}
-C_0^{(G)}\frac{\langle g G^3\rangle}{q^2}+
D_0^{(G)}\frac{\langle \alpha_s^2 G^4\rangle_G}{(q^2)^2},\label{sg}
\end{eqnarray}
where the imaginary part ${\rm Im}\Pi_G^{\rm pert}(s)$ collects the contributions
in Eq.~(\ref{dig3}) without poles at $q^2\to 0$ \cite{Li:2021gsx}, 
\begin{eqnarray}
{\rm Im}\Pi^{\rm pert}_G(s)=-\pi\left[A_0^{(G)}s^2+2A_1^{(G)}s^2\ln\frac{s}{\mu^2}
+A_2^{(G)}s^2\left(3\ln^2\frac{s}{\mu^2}-\pi^2\right)
+B_1^{(G)}\langle \alpha_s G^2\rangle\right].\label{per}
\end{eqnarray}
The term $B_0^{(G)}\langle \alpha_s G^2\rangle$ in Eq.~(\ref{dig3}) does not appear
in the above expression, for it has no discontinuity along the branch cut. 
We work on the subtracted spectral density
\begin{eqnarray}
\Delta\rho_G(s,\Lambda)&=&\rho_G(s)+s^2\left[A_0^{(G)}+2A_1^{(G)}\ln\frac{s}{\mu^2}
+A_2^{(G)}\left(3\ln^2\frac{s}{\mu^2}-\pi^2\right)\right][1-\exp(-s/\Lambda)]\nonumber\\
& &+B_1^{(G)}\langle \alpha_s G^2\rangle[1-\exp(-s^2/\Lambda^2)],\label{co1}
\end{eqnarray}
with $\rho_G(s)\equiv {\rm Im}\Pi_G(s)/\pi$.
The smooth function $1-\exp(-s/\Lambda)$ diminishes like $s$ and $1-\exp(-s^2/\Lambda^2)$ 
diminishes like $s^2$ as $s\to 0$, so the subtraction terms do not modify the 
behavior of $\rho_G(s)\sim s^2$ in the low-energy limit \cite{NSVZ,SVZ80,NS81,MS81}.
Both the smooth functions approach unity at large $s\gg\Lambda$, where the subtracted 
spectral density $\Delta\rho(s,\Lambda)$ decreases quickly, and the radius $R$ can be pushed 
toward the infinity.

Equation~(\ref{sg}) is then converted, in terms of the subtracted spectral density, into 
\begin{eqnarray}
\int_{0}^\infty dy\frac{\Delta\rho_G(y)}{x-y}&=&
-\int_{0}^\infty dy \frac{y^2 e^{-y}}{x-y}\left[A_0^{(G)}+2A_1^{(G)}\ln y
+A_2^{(G)}\left(3\ln^2y-\pi^2\right)\right]\nonumber\\
& &-\int_{0}^\infty dy \frac{ e^{-y^2}}{x-y}B_1^{(G)}\frac{\langle \alpha_s G^2\rangle}{\Lambda^2}
+C_0^{(G)}\frac{\langle g G^3\rangle}{x\Lambda^3}-
D_0^{(G)}\frac{\langle \alpha_s^2 G^4\rangle_G}{x^2\Lambda^4},\label{rg21}
\end{eqnarray}
where the variable changes $x= q^2/\Lambda$ and $y= s/\Lambda$ have been applied, 
the renormalization scale has been set to $\mu^2=\Lambda$, and the dimensionless function 
$\Delta\rho_G(s,\Lambda)/\Lambda^2$ has been expressed as $\Delta\rho_G(y)$ according to 
the argument in the previous section. A stability window of $\Lambda$ is expected 
to emerge, in which a glueball mass $m_G$, read off from a peak location of 
$\Delta\rho_G(s,\Lambda)$, is stable against the variation of $\Lambda$. As $\Lambda$ 
turns sufficiently large, it suppresses the nonperturbative condensate contributions, and 
the scaling of a solution with $\Lambda$ is triggered. When the scaling takes place, no structure 
of a solution can be interpreted as a physical state.

Since the subtracted spectral density behaves as $\Delta\rho_G(y)\sim y^2$ 
at $y\to 0$ as evinced by Eq.~(\ref{co1}), and as $\Delta\rho_G(y)\to 0$ at $y\to\infty$, we 
expand it using the generalized Laguerre polynomials $L_n^{(2)}(y)$ with the index $\alpha=2$ 
in Eq.~(\ref{r1}). Equation~(\ref{ep1}) is inserted into the left-hand side of Eq.~(\ref{rg21}) 
to compute the matrix elements $M_{mn}$ in Eq.~(\ref{m2}), and inserted into the right-hand 
side of Eq.~(\ref{rg21}) to get the coefficients $b_n$ as the inputs. The coefficients $b_1$ 
and $b_2$ of the $1/x$ and $1/x^2$ terms, respectively, receive additional contributions from 
the condensates. To proceed numerical analyses, the four-gluon condensates are approximated, 
under the vacuum factorization assumption, by \cite{NSVZ79,BLP} 
\begin{eqnarray}
\langle \alpha_s^2G^4\rangle_S\approx \frac{9}{16}\langle \alpha_sG^2\rangle^2,\;\;\;\;
\langle \alpha_s^2G^4\rangle_P\approx \frac{15}{8}\langle \alpha_sG^2\rangle^2,
\end{eqnarray}
and the gluon condensate $\langle \alpha_s G^2\rangle$ and the strong coupling $\alpha_s$ are 
the same as in Eq.~(\ref{put}). The triple-gluon condensate 
\begin{eqnarray}
\langle g G^3\rangle &=& -1.5\langle\alpha_s G^2\rangle^{3/2},\label{32}
\end{eqnarray}
is employed from the lattice analysis \cite{PV90}. It has be corroborated \cite{Li:2021gsx} 
that the spectral density obtained from the above result respects better the 
low-energy theorem for the correlation function \cite{NSVZ,SVZ80,LS92}, but those from the 
single-instanton estimate \cite{SVZ,NS80,RRY} and from the fit to heavy quark properties 
\cite{SN10} do not.

\subsection{Scalar Glueballs}

\begin{figure}
\begin{center}
\includegraphics[scale=0.28]{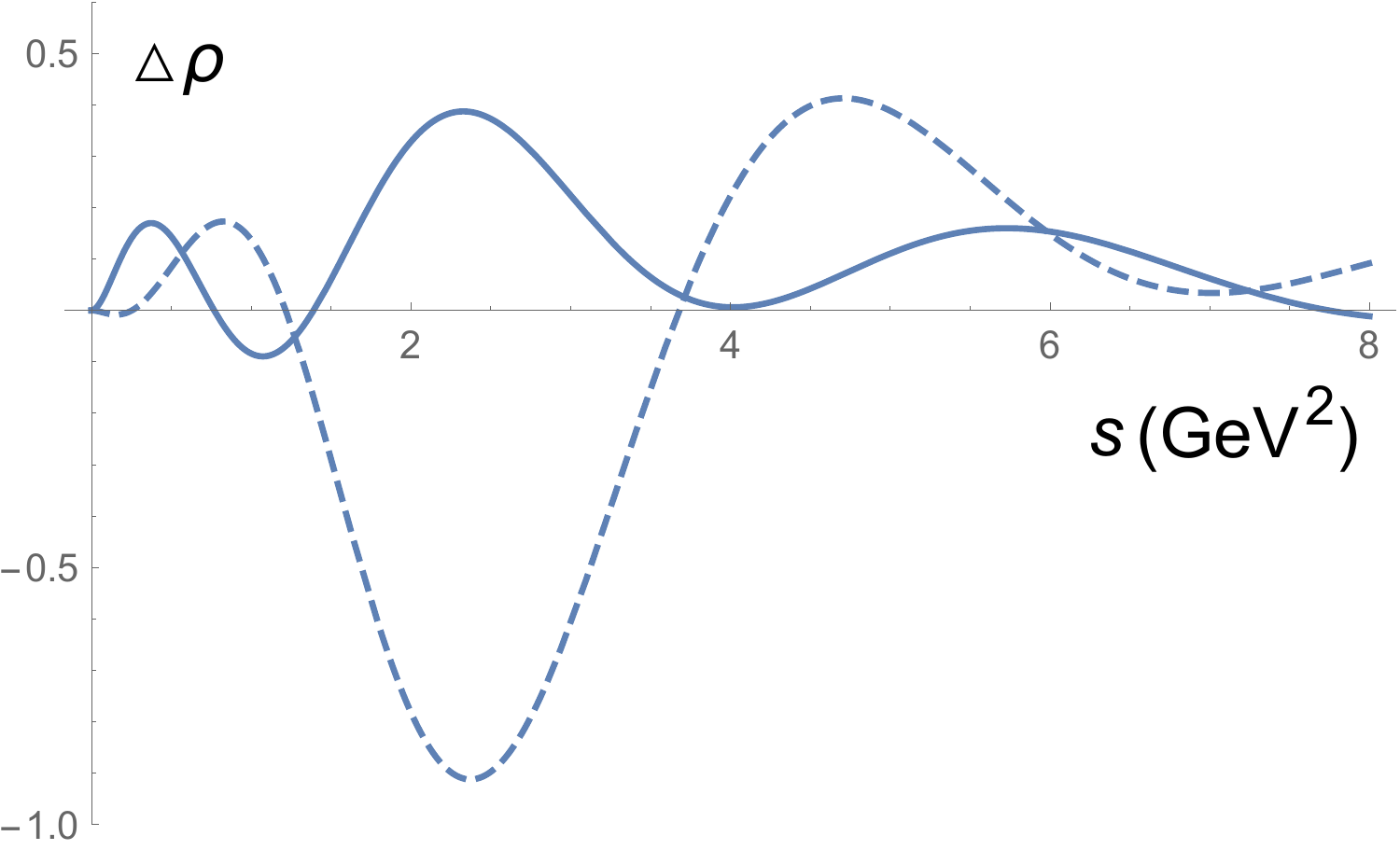}\hspace{1.0cm}
\includegraphics[scale=0.28]{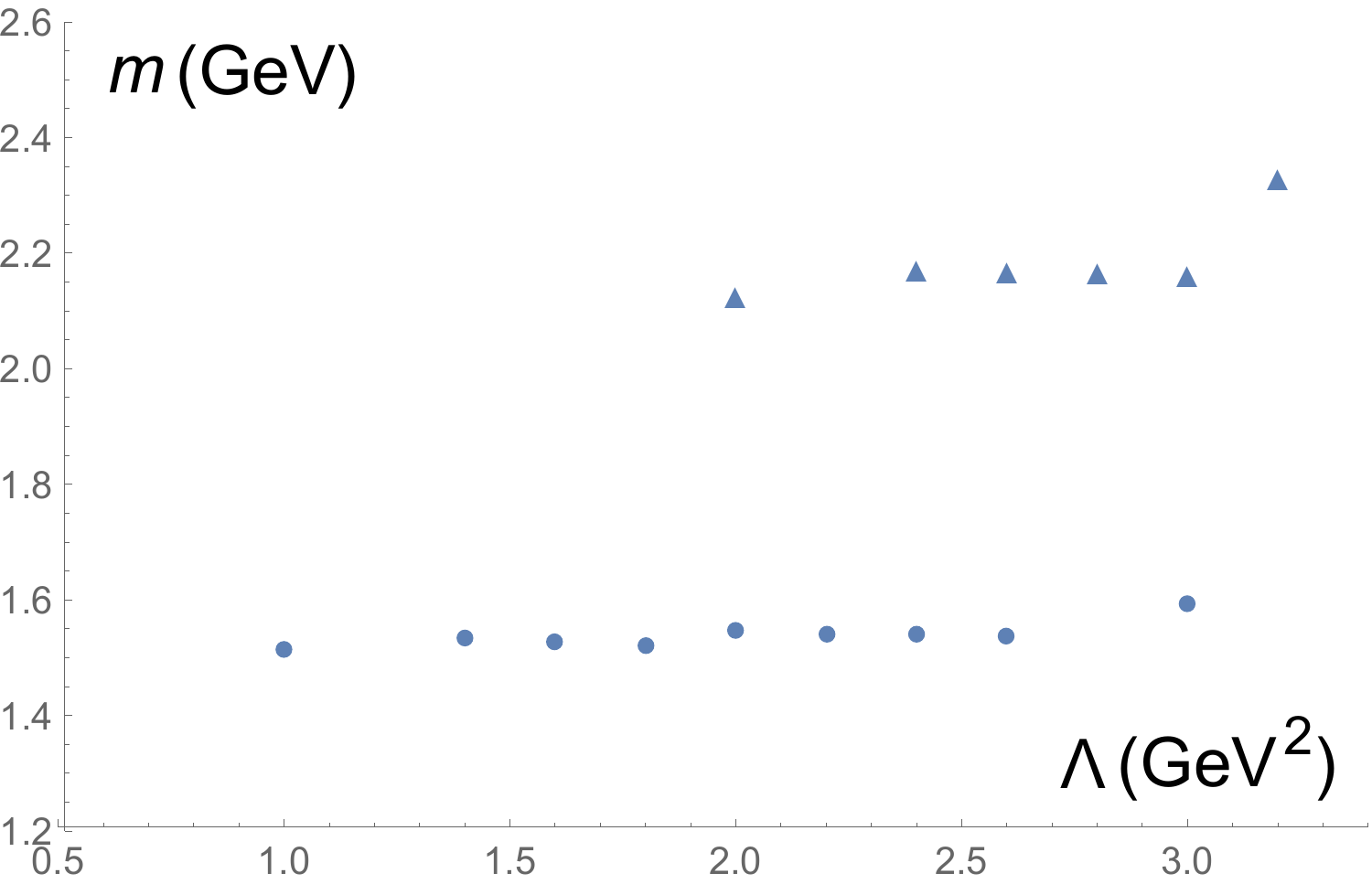}

(a) \hspace{7.0 cm} (b) 
\caption{\label{fig3}
(a) $s$ dependencies of the ground-state solution $\Delta \rho_{S0}(s,\Lambda)$ 
for $\Lambda=1.6$ GeV$^2$ (solid line) and of the first excited-state solution
$\Delta\rho_{S1}(s,\Lambda)$ for $\Lambda=2.8$ GeV$^2$ (dashed line). (b) $\Lambda$ dependencies 
of $m_{S}$ (circles) and $m_{S'}$ (triangles).}
\end{center}
\end{figure}

We solve for the subtracted spectral density $\Delta \rho_{S0}(s,\Lambda)$ associated with
the lightest scalar glueball from Eq.~(\ref{rg21}). The positivity 
constraint on a spectral density is imposed up to the scale $\Lambda=2.8$ GeV$^2$, before 
the coefficient $a_N$ in the polynomial expansion reaches a minimum. Take $\Lambda=1.6$ 
GeV$^2$ as an example, for which $a_{41}$ is the minimum. However, the corresponding spectral 
density $\rho_{S0}(s)$, which is related to $\Delta \rho_{S0}(s,\Lambda)$ via 
Eq.~(\ref{co1}), becomes negative around $s\approx 0.8$ GeV$^2$ as $N=41$. The positivity 
forces the polynomial expansion to terminate at $N=20$ with the ratios of the 
coefficients $|a_{20}/a_1|\approx 0.05$ and $|a_{20}/a_{19}|\approx 1$, which allow 
the polynomial expansion to be under control. The resultant solution 
$\Delta \rho_{S0}(s,\Lambda)$ for $\Lambda=1.6$ GeV$^2$ is shown in Fig.~\ref{fig3}(a). 
It is more practical to work on the subtracted spectral density, instead of the spectral
density, for the huge perturbative contribution washes out the structure of the former
easily. It is interesting to have two peaks, one at $s=0.37$ GeV$^2$, i.e., 
$\sqrt{s}\approx 0.61$ GeV and another at $s=2.34$ GeV$^2$, i.e.,
$\sqrt{s}\approx 1.53$ GeV, similar to the observation in \cite{NV89,Harnett:2000fy}. As 
elucidated in \cite{Li:2021gsx}, the shorter peak located at 0.61 GeV is identified as the 
light scalar meson $f_0(500)$, and the taller peak located at 1.53 GeV corresponds to a
glue-rich state, i.e., the lightest scalar glueball. The mass 1.53 GeV is very close to the 
measured $f_0(1500)$ meson mass $1522\pm 25$ MeV \cite{PDG}. In view of the broad width about 
300 MeV, the taller peak may arise from the combined contribution of the $f_0(1370)$, 
$f_0(1500)$ and $f_0(1700)$ mesons, in that $f_0(1500)$ with a narrow width $108\pm 33$ MeV 
\cite{PDG} cannot accommodate the profile alone. This observation matches the prevailing 
consensus in the literature 
\cite{Close,Giacosa:2005zt,Vento:2004xx,Fariborz:2006xq,Fariborz:2003uj,
Cheng:2015iaa,Noshad:2018afw,Guo:2020akt}; the predicted scalar glueball mass 
$m_S\approx 1.53$ GeV is consistent with those from sum rules 
\cite{SN98,Forkel:2000fd,Wen:2010qoe,Narison:2021xhc} 
(but a bit lower than in \cite{Chen:2021bck}) and quenched lattice QCD 
\cite{Bali,Morningstar:1999rf,Chen:2005mg,Athenodorou:2020ani}, but
smaller than from holographic QCD \cite{Zhang:2021itx}.

We survey the variation of the ground-state mass $m_{S}$ with the scale $\Lambda$ in the 
range $1.0 \le\Lambda \le 3.4$ GeV$^2$. The highest degree $N$ for the polynomial expansion 
increases with $\Lambda$ under the positivity requirement, $N=12$, 17, 20, 23, 25, 28, 31 and 
34 for $\Lambda=1.0$, 1.4, 1.6, 1.8, 2.0, 2.2, 2.4 and 2.6 GeV$^2$, respectively. As $\Lambda$ 
goes up further, the positivity ceases to function, and $N=39$ and 38 are selected 
for $\Lambda=3.0$ and 3.4 GeV$^2$, respectively, corresponding to the minimal $a_N$. The 
$\Lambda$ dependence of $m_{S}$ is displayed in Fig.~\ref{fig3}(b), where the curve ascends 
from $\Lambda=1.0$ GeV$^2$, reaches $m_{S}\approx 1.53$ GeV, keeps more or less flat 
in the window $1.4\le \Lambda \le 2.6$ GeV$^2$, and then rises again monotonically. The 
behavior in the region with $\Lambda>2.6$ GeV manifests the scaling of the solutions 
ascribed to the disappearance of the nonperturbative condensate effects. We assess 
the uncertainty in our method by means of the fluctuation in the stability 
window, arriving at $m_{S}=(1.53\pm 0.02)$ GeV, whose tiny errors reflect the remarkable 
rigidity of our solutions. The uncertainties from the other sources have been examined in 
\cite{Li:2021gsx}, and concluded to be moderate; the $\pm 20\%$ change 
of the gluon condensate $\langle\alpha_s G^2\rangle$ causes about $\pm 5\%$ impact.
We thus have the prediction $m_S=(1.53\pm 0.07)$ GeV, combining the above two sources of
theoretical uncertainties.

As in Sec.~IID, we parametrize the ground-state contribution to the
spectral density as a $\delta$-function $F_{S}\delta(s-m_S^2)$, and subtract it from the two 
sides of the dispersion relation in Eq.~(\ref{rg21}). The strength $F_S$ of the $\delta$-function 
is calculated from the ground-state solution $\Delta\rho_{S0}(s,\Lambda)$ corresponding to 
the high end of the stability window, i.e., $\Lambda=2.6$ GeV$^2$, 
\begin{eqnarray}
F_{S}=\int_0^\infty ds\Delta\rho_{S0}(s,\Lambda)=1.50\;{\rm GeV}^2.
\end{eqnarray}
The dispersion relation associated with the first excited scalar glueball
then reads
\begin{eqnarray}
\int_{0}^\infty dy\frac{\Delta\rho_S(y)}{x-y}&=&
-\int_{0}^\infty dy \frac{y^2 e^{-y}}{x-y}\left[A_0^{(S)}+2A_1^{(S)}\ln y
+A_2^{(S)}\left(3\ln^2y-\pi^2\right)\right]\nonumber\\
& &-\int_{0}^\infty dy \frac{ e^{-y^2}}{x-y}B_1^{(S)}\frac{\langle \alpha_s G^2\rangle}{\Lambda^2}
-\int_{0}^\infty dy \frac{f_{S}\delta(y-r_{S})}{x-y}
+C_0^{(S)}\frac{\langle g G^3\rangle}{x\Lambda^3}-
D_0^{(S)}\frac{\langle \alpha_s^2 G^4\rangle_S}{x^2\Lambda^4},\label{rg22}
\end{eqnarray}
with the ratios $f_{S}=F_{S}/\Lambda$ and $r_{S}=m_S^2/\Lambda$. The inputs $b_n$ arise from 
the right-hand side of Eq.~(\ref{rg22}) with the same OPE parameters in Eqs.~(\ref{put}) and
(\ref{32}).

In this case the coefficient $a_N$ reaches a minimum before the positivity of the spectral 
density is violated. The $s$ dependence of the solution $\Delta\rho_{S1}(s,\Lambda)$ for 
$\Lambda=2.8$ GeV$^2$ with the highest degree $N=37$ is presented in Fig.~\ref{fig3}(a). 
The deep valley at $s\approx 2.3$ GeV$^2$, i.e., at $m_S\approx 1.5$ GeV$^2$ is due to the 
subtraction of the ground-state contribution. Figure~\ref{fig3}(a) contrasts the valley in 
the first excited-state solution with the peak in the ground-state solution. The oscillation 
above $s\approx 6.0$ GeV$^2$ is similar to those unveiled before. The prominent peak located 
at $s=4.71$ GeV$^2$, i.e., at the mass $m_{S'}=2.17$ GeV will be identified as the first 
excited scalar glueball. We tend to assign the $f_0(2200)$ meson as the candidate, because 
of its mass $2187\pm 14$ MeV close to the peak location and the $O(10^{-4})$ branching ratio 
for the radiative decay $J/\psi\to \gamma f_0(2200)$ \cite{PDG}. Note that the peak width is 
broad, also covering the $f_0(2100)$ and $f_0(2330)$ mesons, which have been considered 
as possible candidates for excited scalar glueballs in a phenomenological 
study \cite{Sonnenschein:2018fph}. Their production rates in the $J/\psi$ radiative decays 
are also of $O(10^{-4})$ \cite{PDG}. Since the nearby $f_0(2100)$, $f_0(2200)$ and $f_0(2330)$ 
states may not be resolved in our framework, the prominent peak in Fig.~\ref{fig3}(a) is 
likely to be an admixture of these three resonances.

\begin{figure}
\begin{center}
\includegraphics[scale=0.28]{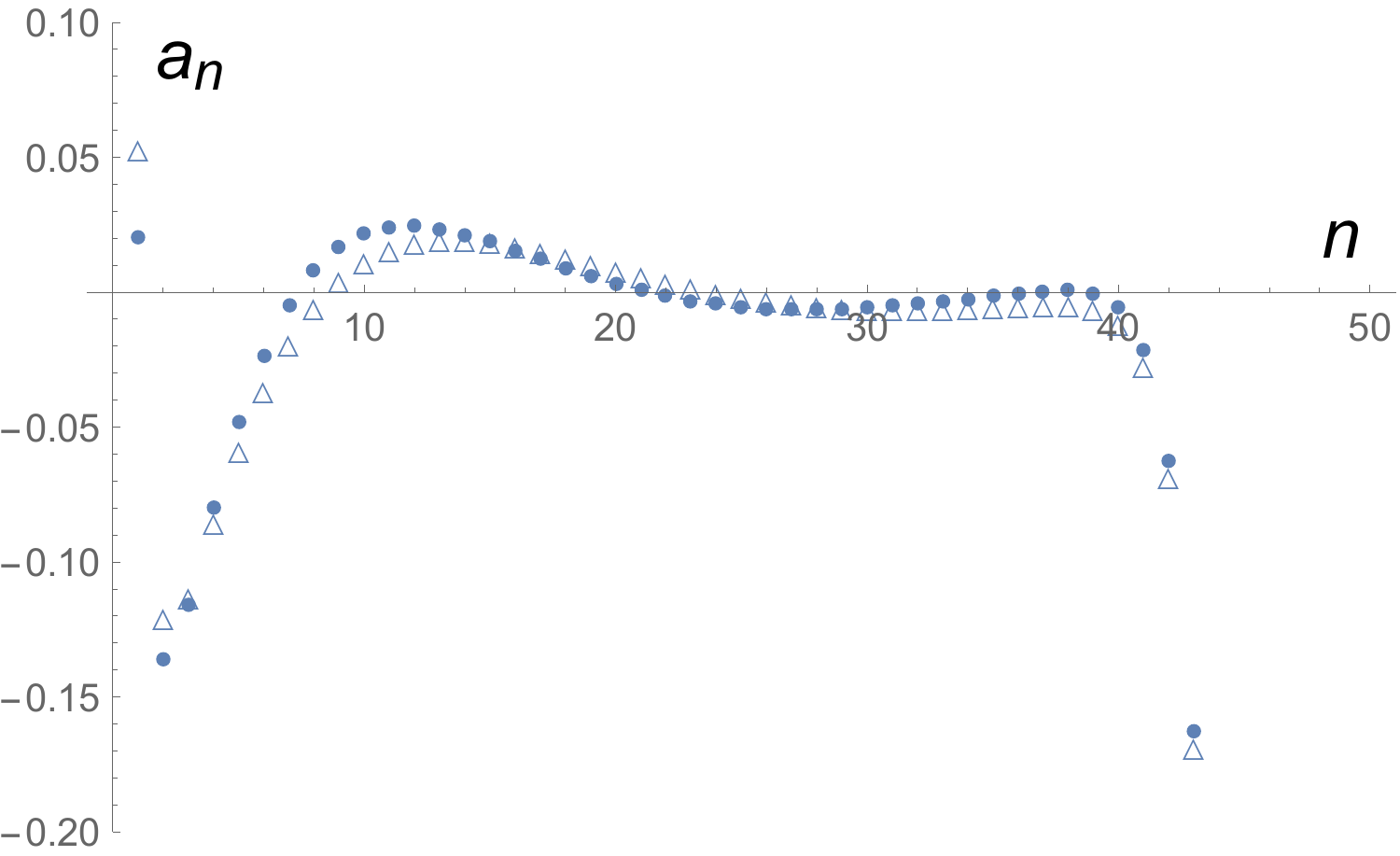}

\caption{\label{fig4}
Solved coefficients $a_n$ for $\Lambda=2.8$ 
GeV$^2$ (filled circles) and $\Lambda=3.2$ GeV$^2$ (empty circles).}
\end{center}
\end{figure}

We then explore the $\Lambda$ dependence of the first excited scalar glueball mass $m_{S'}$ 
by scanning the interval $2.0\le \Lambda\le 3.2$ GeV$^2$. The maximal degree $N$ takes the 
integers 32, 31, 34, 37, 38 and 24 for $\Lambda=2.0$, 2.4, 2.6, 2.8, 3.0 and 3.2 GeV$^2$, 
respectively. As $\Lambda=2.0$ GeV$^2$, the positivity requirement is effective, similar to 
what happened to the case of the lightest scalar glueball. The big drop in $N$ at 
$\Lambda=3.2$ GeV can be explained by means of Fig.~\ref{fig4}, which compares the solved 
coefficients $a_n$ for $\Lambda=2.8$ and 3.2 GeV$^2$. The sudden descent near $n=40$ 
is a consequence of the ill-posed nature of an inverse problem. As stated before, a 
large $N$ is preferred in order to suppress the $1/(q^2)^{N+1}$ power correction to a 
dispersion relation. It is apparent that $a_{37}$ is the minimum in the $\Lambda=2.8$ case, 
which, however, turns negative in the $\Lambda=3.2$ case. As seen below, the turning signals 
the ending of the stability window. One then has to choose $N=24$, which points to a minimal 
$a_N$ in the low $n$ region. The curve for $m_{S'}$ in Fig.~\ref{fig3}(b) climbs 
from $\Lambda=2.0$ GeV, enters the stability window $2.4\le \Lambda\le 3.0$ GeV$^2$, and 
ascends quickly as $\Lambda>3.0$ GeV$^2$. The stability window for the first excited state 
is also narrower than for the ground state, the same as in Fig.~\ref{fig2}(b) for the series of
$\rho$ resonances. It is straightforward to retrieve $m_{S'}=2.17\pm 0.01$ GeV from the 
stability window in agreement with the measured $f_0(2200)$ meson mass \cite{PDG}. The 
representative 20\% variation of the gluon condensate induces about 1\% effect, which is 
negligible. Another choice of the ground-state solution $\Delta\rho_{S0}(s,\Lambda)$ 
corresponding to a smaller $\Lambda=2.4$ GeV$^2$, which yields the strength $F_S=1.39$ GeV$^2$, 
does not alter the result.

We test the analysis on the second excited scalar glueball, though a meaningful outcome may 
not be expected as conjectured in the previous section. The profile of the $f_0(2200)$ 
contribution to the spectral density is parametrized as a $\delta$ function 
$F_{S'}\delta(s-m_{S'}^2)$. The solution corresponding to the high end of the stability 
window, i.e., $\Lambda=2.8$ GeV$^2$ with $N=37$, is selected to compute the strength
\begin{eqnarray}
F_{S'}=\int_{t_1}^\infty ds\Delta\rho_{S1}(s,\Lambda)=1.37\;{\rm GeV}^2.
\end{eqnarray}
The lower bound $t_1=3.68$ GeV$^2$ is set to the first root in $s>0$ of the equation
$\Delta\rho_{S1}(s,\Lambda)=0$ to delete the contribution from the deep valley. 
The dispersion relation for the second excited scalar glueball is then given by
\begin{eqnarray}
\int_{0}^\infty dy\frac{\Delta\rho_S(y)}{x-y}&=&
-\int_{0}^\infty dy \frac{y^2 e^{-y}}{x-y}\left[A_0^{(S)}+2A_1^{(S)}\ln y
+A_2^{(S)}\left(3\ln^2y-\pi^2\right)\right]\nonumber\\
& &-\int_{0}^\infty dy \frac{ e^{-y^2}}{x-y}B_1^{(S)}\frac{\langle \alpha_s G^2\rangle}{\Lambda^2}
-\int_{0}^\infty dy \frac{f_{S}\delta(y-r_{S})+f_{S'}\delta(y-r_{S'})}{x-y}\nonumber\\
& &+C_0^{(S)}\frac{\langle g G^3\rangle}{x\Lambda^3}-
D_0^{(S)}\frac{\langle \alpha_s^2 G^4\rangle_S}{x^2\Lambda^4},\label{rg23}
\end{eqnarray}
with the ratios $f_{S'}=F_{S'}/\Lambda$ and $r_{S'}=m_{S'}^2/\Lambda$. The same steps lead
to the mass $m_{S''}\approx 2.84$ GeV for the second excited scalar glueball. It is
not sure which known state it should be assigned to in view of the potential sizable
theoretical error. However, the result conforms the typical energy gap of about 700 MeV 
between two neighboring states generated by radial excitations.

\subsection{Pseudoscalar Glueballs}

\begin{figure}
\begin{center}
\includegraphics[scale=0.28]{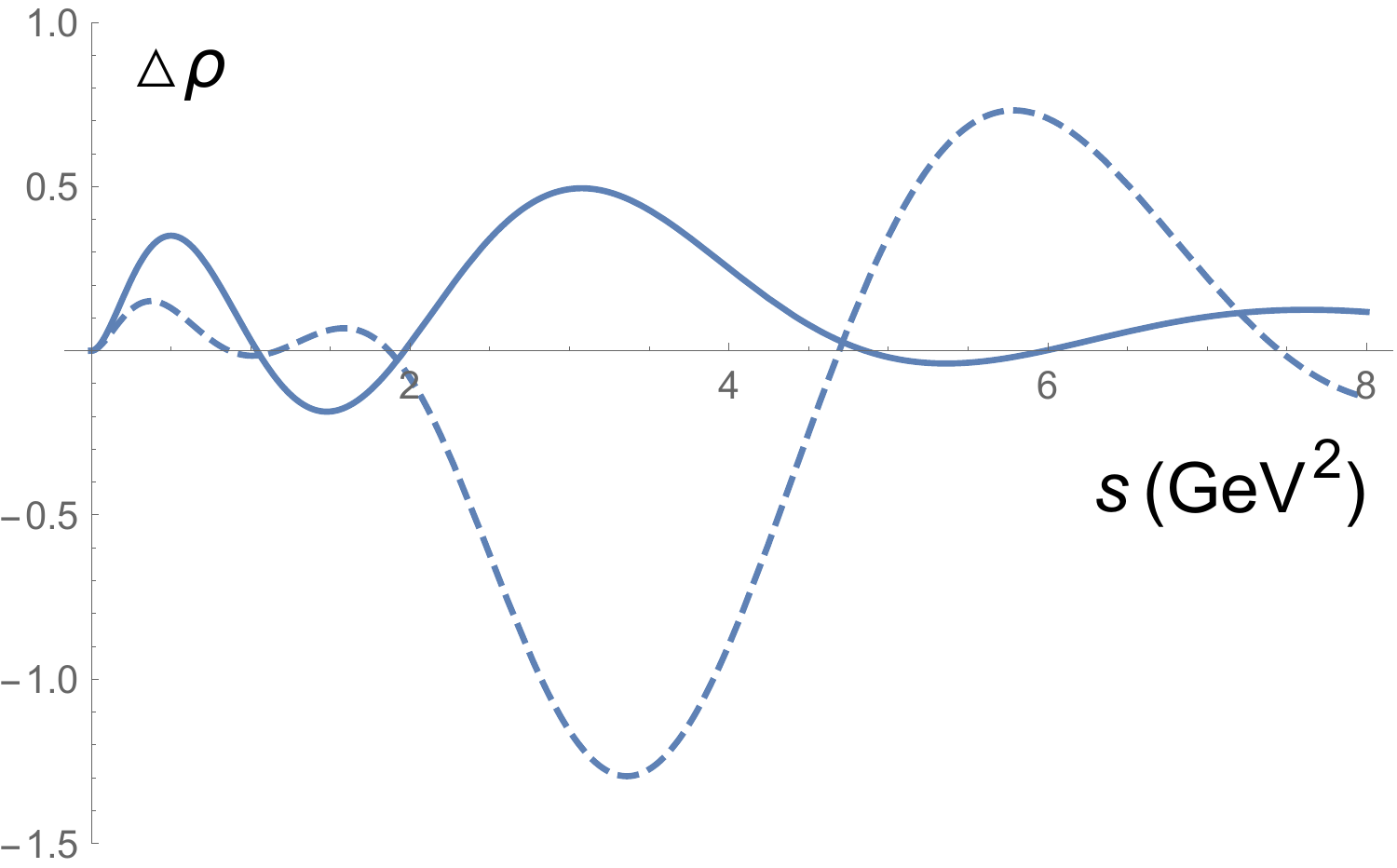}\hspace{1.0cm}
\includegraphics[scale=0.28]{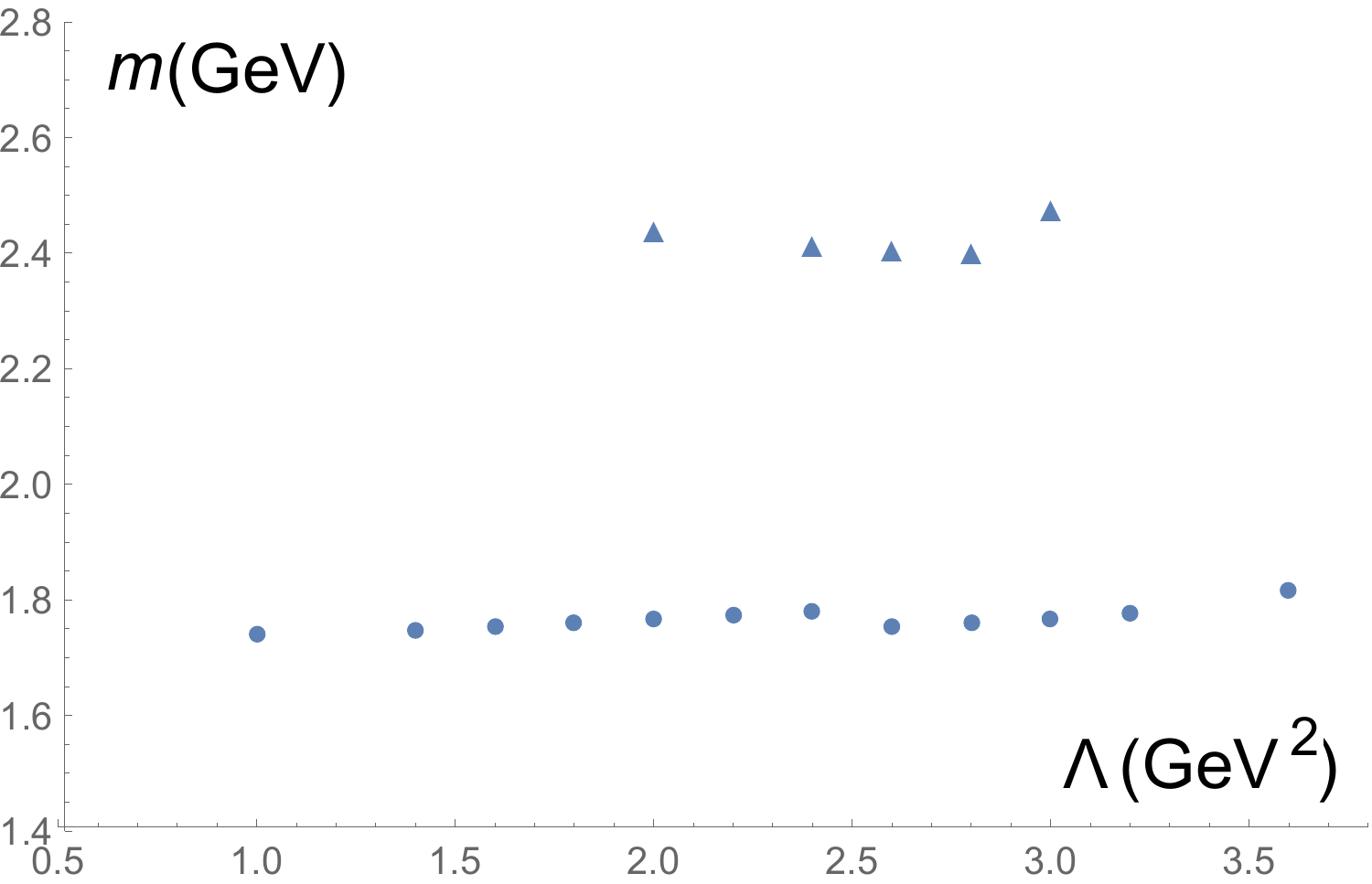}

(a) \hspace{7.0 cm} (b)
\caption{\label{fig5}
(a) $s$ dependencies of the ground-state solution $\Delta \rho_{P0}(s,\Lambda)$ 
for $\Lambda=1.6$ GeV$^2$ (solid line) and of the first excited-state solution
$\Delta\rho_{P1}(s,\Lambda)$ for $\Lambda=2.8$ GeV$^2$ (dashed line). (b) $\Lambda$ dependencies 
of $m_{P}$ (circles) and $m_{P'}$ (triangles).}
\end{center}
\end{figure}

At last, we determine the lightest and excited pseudoscalar glueball masses by solving 
Eq.~(\ref{rg21}) from the OPE inputs in Eq.~(\ref{ap}) and the triple-gluon condensate in 
Eq.~(\ref{32}). The prescription for fixing the highest degree $N$ for the polynomial 
expansion is the same; we look for the minimal $a_N$ before the positivity constraint is 
activated. The quality of the expansion is similar to the scalar glueball case. 
The behavior of the ground-state solution 
$\Delta\rho_{P0}(s,\Lambda)$ for the subtracted spectral density at $\Lambda=1.6$ GeV$^2$ with 
$N=15$ is plotted in Fig.~\ref{fig5}(a). Likewise, we observe the double-peak structure with
oscillations at large $s$. The shorter peak located at $s=0.50$ GeV$^2$, i.e., $\sqrt{s}=0.71$ GeV 
comes from the combined contribution of the $\eta$ and $\eta'$ mesons, which have been known to 
comprise some gluonium components \cite{Kataev:1981aw}. Such a low-lying state with mass around 
1 GeV was also identified and assigned to the $\eta'$ meson in the lattice calculation 
\cite{Sun:2017ipk}. The major peak of width about 270 MeV located at $s=3.06$ GeV$^2$, i.e., 
$\sqrt{s}=1.75$ GeV fits the $\eta(1760)$ meson \cite{Li:2021gsx} with the mass $1751\pm 15$ MeV 
and width $240\pm 30$ MeV \cite{PDG}. This mass is just a bit larger than the scalar 
glueball one derived in the previous subsection, as anticipated from the minor difference 
between their OPE inputs and argued in \cite{Faddeev:2003aw}. 

The ground-state mass $m_P=1.75$ GeV for the pseudoscalar glueball is lower than most findings 
above 2 GeV in the literature, such as those from sum rules \cite{SN98}, quenched lattice QCD 
\cite{Bali,Morningstar:1999rf,Chen:2005mg,Athenodorou:2020ani}, the Bethe-Salpeter approach 
\cite{Huber:2020ngt,Kaptari:2020qlt} and holographic QCD \cite{Zhang:2021itx}. Nevertheless, 
when the resonance contribution was parametrized by a Breit-Wigner form with a finite width, 
the pseudoscalar glueball mass drops to $(1.407\pm 0.162)$ GeV in sum rules with the instanton 
effect \cite{Wang:2015mla}. Our solution is heavier than the $\eta(1405)$ meson, which was 
speculated to be the lightest pseudoscalar glubeball \cite{MCU} in a pseudoscalar 
meson mixing formalism based on the anomalous Ward identity \cite{Cheng:2008ss,He:2009sb,
Tsai:2011dp}. It should be pointed out, however, that a pseudoscalar glueball mass 
as high as 1.75 GeV is not excluded in \cite{Cheng:2008ss}, when some inputs are tuned 
\cite{Qin:2017qes}; our prediction is consistent with the results in \cite{Qin:2017qes}
with a large angle $\phi_G$ for the mixing between the pure glueball and the flavor-singlet 
light quark states. We advocate that the $\eta(1760)$ meson is a promising candidate for the 
lightest pseudosclar glueball, which was proposed almost three decades ago 
\cite{Page:1996ss,Wu:2000yt}, and examined experimentally via the decay 
$J/\psi\to\gamma(\eta(1760)\to)\omega\omega$ in \cite{BES:2006nqh}. The proposal in 
\cite{Page:1996ss} was motivated by the large branching ratio of $\eta(1760)$ decays into 
gluon pairs, which was extracted from the data of $J/\psi$ and $\Upsilon$ radiative decays. 
The proposal in \cite{Wu:2000yt} was based on the mass relation for the mixing among the 
pseudoscalar glueball and mesons, which reduced the higher pure glueball mass to the lower 
mixed glueball mass around 1.7 GeV. $\eta(1760)$ mesons are abundantly produced 
in $J/\psi$ radiative decays with the branching ratio of $O(10^{-3})$, but not seen in the 
$J/\psi\to \gamma\gamma V$ channels \cite{MARK}, $V=\rho$, $\phi$, implying that the partial 
width of $\eta(1760)\to \gamma V$ is tiny. Another nearby pseudoscalar $X(1835)$ is produced 
less abundantly in $J/\psi$ radiative decays, but seen in the $J/\psi\to \gamma\gamma \phi$ 
decay \cite{BESIII:2018dim}. 

We find the ground-state masses $m_{P}$ from the solutions for the subtracted spectral density 
in the range $1.0\le \Lambda \le 3.6$ GeV$^2$. The maximal degree $N$ increases with $\Lambda$ 
under the positivity constraint, taking $N=9$, 13, 15, 17, 19, 21, 23, 26, 28, 30, 32 and 36 
for $\Lambda=1.0$, 1.4, 1.6, 1.8, 2.0, 2.2, 2.4, 2.6, 2.8, 3.0 3.2 and 3.6 GeV$^2$, respectively. Figure~\ref{fig5}(b) describes the dependence of $m_P$ on $\Lambda$, in which the 
curve ascends from $\Lambda=1.0$ GeV$^2$, almost keep flat in the interval 
$1.4\le \Lambda\le 3.2$ GeV$^2$, and goes up monotonically as $\Lambda>3.2$ GeV$^2$. The 
stability window is wider than in Fig.~\ref{fig3}(b) for the lightest scalar glueball. We 
estimate the theoretical uncertainty in our method from the fluctuation within the stability 
window, and put forth $m_{P}=(1.75\pm 0.02)$ GeV, whose small errors reflect the stability of our 
solutions. The $\pm 20\%$ change of the gluon condensate stimulates $\pm 5\%$ effect. It is 
apparent that our prediction $m_P=1.75\pm 0.10$ GeV, whose errors combine the above two sources, 
coincides with the measured mass of the $\eta(1760)$ meson \cite{PDG}.

Motivated by the recent affirmation of the $X(2370)$ quantum numbers as $0^{-+}$, we 
extract the excited pseudoscalar glueball mass in our framework. The ground-state 
contribution to the spectral density is parametrized as a $\delta$-function
$F_{P}\delta(s-m_{P}^2)$, and subtracted from the two sides of the dispersion relation in 
Eq.~(\ref{rg21}). The ground-state solution $\Delta\rho_{P0}(s,\Lambda)$ corresponding to the 
high end of the stability window, i.e., $\Lambda=3.2$ GeV$^2$ with $N=32$, is 
picked up for the evaluation of the strength
\begin{eqnarray}
F_{P}=\int_0^\infty ds\Delta\rho_{P0}(s,\Lambda)=2.33\;{\rm GeV}^2.
\end{eqnarray}
Adjusting the strength $F_{P}$ to those from the other ground-state solutions corresponding to, 
say, $\Lambda=3.0$ or 3.4 GeV$^2$, we observe less than 1\%, i.e., negligible impact on the
results. The dispersion relation for the first excited pseudoscalar glueball is given by
\begin{eqnarray}
\int_{0}^\infty dy\frac{\Delta\rho_P(y)}{x-y}&=&
-\int_{0}^\infty dy \frac{y^2 e^{-y}}{x-y}\left[A_0^{(P)}+2A_1^{(P)}\ln y
+A_2^{(P)}\left(3\ln^2y-\pi^2\right)\right]\nonumber\\
& &-\int_{0}^\infty dy \frac{ e^{-y^2}}{x-y}B_1^{(P)}\frac{\langle \alpha_s G^2\rangle}{\Lambda^2}
-\int_{0}^\infty dy \frac{f_{P}\delta(y-r_{P})}{x-y}
+C_0^{(P)}\frac{\langle g G^3\rangle}{x\Lambda^3}-
D_0^{(P)}\frac{\langle \alpha_s^2 G^4\rangle_P}{x^2\Lambda^4},\label{rg24}
\end{eqnarray}
with the ratios $f_{P}=F_{P}/\Lambda$ and $r_{P}=m_P^2/\Lambda$. The inputs $b_n$ are prepared
according to the right-hand side of Eq.~(\ref{rg24}) using the OPE parameters in 
Eqs.~(\ref{put}) and (\ref{32}).

The coefficients $a_n$ obtained from Eq.~(\ref{rg24}) converge more slowly, so the positivity 
of the spectral density forces the termination of the polynomial expansion. The $s$ dependence 
of the solution $\Delta\rho_{P1}(s,\Lambda)$ for $\Lambda=2.8$ GeV$^2$ with the maximal degree 
$N=36$ is depicted in Fig.~\ref{fig5}(a). The deep valley around $s=3.3$ GeV$^2$ traces back to 
the subtraction of the ground-state contribution, as contrasted by the peak associated with the 
$\eta(1760)$ meson. The peak located at $s= 5.81$ GeV$^2$ next to the valley, i.e., at the mass 
$m_{P'}=2.41$ GeV of the first excited pseudoscalar glueball will be identified as the 
$X(2370)$ meson. We investigate the $\Lambda$ dependence of $m_{P'}$ by scanning the interval 
$2.0\le \Lambda\le 3.0$ GeV$^2$. The curve in Fig.~\ref{fig5}(b) descends gradually from 
$\Lambda=2.0$ GeV$^2$, reaches minima in the region $2.4\le\Lambda\le 2.8$ GeV$^2$, and then 
rises as $\Lambda>2.8$ GeV$^2$, manifesting the scaling behavior. We designate the region 
$2.4\le \Lambda\le 2.8$ as the stability window, which is also narrower than for the ground 
state. The highest degree $N$ reads 24, 30, 33, 36 and 29 for $\Lambda=2.0$, 2.4, 2.6, 2.8 and 
3.0 GeV$^2$, respectively. The reason for the drop of $N$ at $\Lambda=3.0$ GeV$^2$ is the same, 
which has been illustrated in Fig.~\ref{fig4} for the first excited scalar glueball. 
The $\pm 20\%$ variation of the gluon condensate induces about $\pm 2\%$ change. 
We summarize our prediction for the mass of the first excited pseudoscalar glueball, 
$m_{P'}=2.41\pm 0.04$ GeV, consistent with the BESIII measurement 
$2395\pm 11({\rm stat})^{+26}_{-94}({\rm syst})$ MeV \cite{BESIII:2023wfi}. 
The branching ratio of the radiative decay 
$J/\psi\to \gamma X(2370)$, estimated to be of $O(10^{-4})$, is similar to 
${\rm Br}(J/\psi\to \gamma f_0(2200))$ in the scalar glueball case.

We also check the possibility to probe the second excited pseudoscalar glueball, 
parametrizing the profile of the $X(2370)$ contribution to the spectral density as a 
$\delta$-function $F_{P'}\delta(s-m_{P'}^2)$. The solution corresponding to the 
high end of the stability window, i.e., $\Lambda=2.8$ GeV$^2$ with $N=36$, is employed  
for the evaluation of the strength,
\begin{eqnarray}
F_{P'}=\int_{t1}^\infty ds\Delta\rho_{P1}(s,\Lambda)=1.74\;{\rm GeV}^2.
\end{eqnarray}
The lower bound $t_1=4.69$ GeV$^2$ is fixed by the first root in $s>0$ of the equation
$\Delta\rho_{P1}(s,\Lambda)=0$. The dispersion relation for the second excited scalar glueball 
is expressed as
\begin{eqnarray}
\int_{0}^\infty dy\frac{\Delta\rho_S(y)}{x-y}&=&
-\int_{0}^\infty dy \frac{y^2 e^{-y}}{x-y}\left[A_0^{(P)}+2A_1^{(P)}\ln y
+A_2^{(P)}\left(3\ln^2y-\pi^2\right)\right]\nonumber\\
& &-\int_{0}^\infty dy \frac{ e^{-y^2}}{x-y}B_1^{(P)}\frac{\langle \alpha_s G^2\rangle}{\Lambda^2}
-\int_{0}^\infty dy \frac{f_{P}\delta(y-r_{P})+f_{P'}\delta(y-r_{P'})}{x-y}\nonumber\\
& &+C_0^{(P)}\frac{\langle g G^3\rangle}{x\Lambda^3}-
D_0^{(P)}\frac{\langle \alpha_s^2 G^4\rangle_P}{x^2\Lambda^4},\label{rg25}
\end{eqnarray}
with the ratios $f_{P'}=F_{P'}/\Lambda$ and $r_{P'}=m_{P'}^2/\Lambda$. We get the mass 
$m_{P''}\approx 3.21$ GeV for the second excited pseudoscalar glueball, which is above the 
first excited state by a reasonable gap 800 MeV. For a similar reason,  
we are not sure which known state it should be assigned to, considering the potential sizable
theoretical error.

\section{CONCLUSION}

We have refined our proposal for handling QCD sum rules as an inverse problem, i.e., solving 
a dispersion relation with OPE inputs directly. It was shown in our previous study 
that masses of ground states, including the $\rho(770)$ meson, the scalar glueball 
as an admixture of the $f_0(1370)$, $f_0(1500)$ and $f_0(1710)$ mesons, and the pseudoscalar 
glueball $\eta(1760)$, can be determined before the ill-posedness of an inverse problem 
emerges. The present work aims at an extension to excited 
states. The idea is to parametrize a ground-state solution for a subtracted spectral density 
as a $\delta$-function, which is nonvanishing only at the ground-state mass and whose strength 
is set to the integrated ground-state contribution. The purpose of this approximation is to 
reduce the interference between the ground and excited states, since the tail of the 
former is expected to affect the behavior of the latter easily. The $\delta$-function is 
subtracted from two sides of a dispersion relation, which is then solved with the
same OPE inputs again. Stability windows for solutions were found to exist, allowing 
the identification of physical excited states; the $\rho(1450)$ and $\rho(1700)$ resonances 
were constructed with the predicted masses $m_{\rho'}=1.47\pm 0.02$ GeV and 
$m_{\rho''}\approx 1.65$ GeV, respectively. The larger deviation associated with a higher excited 
state was discussed; the uncertainty inherent in a lower state may propagate 
into higher states and be accumulated through the repeated subtractions.

The above strategy, as implemented for the cases of glue-rich states, suggests the $f_0(2200)$ 
and $X(2370)$ mesons to be the first excited scalar and pseudoscalar glueballs, respectively. 
The predicted mass $m_{S'}=2.17\pm 0.01$ GeV for the first excited scalar glueball agrees with 
the observed $f_0(2200)$ meson mass, and the width is broad enough to cover the $f_0(2100)$ and 
$f_0(2330)$ resonances. That is, the solution is likely to be an admixture of the $f_0(2100)$, 
$f_0(2200)$ and $f_0(2330)$ mesons. The predicted mass $m_{P'}=2.41\pm 0.04$ GeV for the 
first excited pseudoscalar glueball is close to the BESIII measurement for the $X(2370)$ meson. 
As stressed in the Introduction, it is difficult to access pseudoscalar glueballs by quenched 
lattice QCD owing to the involved axial anomaly. Therefore, assigning $X(2370)$ to the lightest 
pseudoscalar glueball simply based on the similar mass range 2.3-2.6 GeV from quenched 
lattice QCD might be in doubt. We would rather identify $\eta(1760)$ as the lightest 
pseudoscalar glueball, and $X(2370)$ as its excited state; the simultaneous accommodation 
of the consensual spectra for the $\rho$ and scalar glueball resonances in our dispersive 
analyses strongly supports this postulation. Both $f_0(2200)$ and $X(2370)$ are heavier 
than their ground states by about 700 MeV, a typical energy gap induced by radial 
excitations. The $O(10^{-4})$ branching ratios of the radiative decays 
$J/\psi\to\gamma f_0(2200)$ and $\gamma X(2370)$, lower than those of $O(10^{-3})$ associated 
with the ground states, also make sense.  

The current framework is worth improving for more precise pursuits of higher excited states. 
Sophisticated parametrization and subtraction of lower-state contributions to 
spectral densities need to be developed. The interference effects among nearby resonances 
should be taken into account, as remarked in Sec.~IID. We will address these issues in future 
publications.

\vskip 1.0cm
{\bf Acknowledgement}

We thank F.K. Guo, S. Jin and Z.X. Zhao for stimulating discussions. 
%This work was supported in part by the Ministry of Science and Technology of R.O.C. 
%under Grant No. NSTC-113-2112-M-001-024-MY3.

\end{document}